\documentclass[11pt,a4paper,oneside]{article}
\pdfoutput=1
\usepackage{jheppub}
\usepackage{verbatim}
\usepackage{mathrsfs}
\usepackage{appendix}
\usepackage{mathtools}
\usepackage{float}
\usepackage{subcaption}
\usepackage{alltt}
\usepackage{array,multirow}
\usepackage{lscape}
\usepackage{graphicx}
\usepackage{amssymb,amsmath,amsfonts,mathrsfs,enumerate,latexsym,wasysym}

\usepackage{graphicx}
\usepackage{adjustbox}
\usepackage{xcolor}  
\usepackage{slashed}    
\usepackage{url}
\usepackage{hyperref}
\usepackage[numbers,sort&compress]{natbib}
\usepackage{fancyvrb}
\allowdisplaybreaks
\usepackage{multirow}
\usepackage{color}
\usepackage{pifont}
\usepackage{filecontents}
\usepackage{mathtools}
\usepackage{setspace}
\usepackage{lscape}

\title{\bf Memories of quenches in operator mixing}

\author{\sf Joydeep Chakrabortty, Diptarka Das, Bidyut Dey, Suraj Prakash, and Shakeel Ur Rahaman}

\emailAdd{joydeep@iitk.ac.in, didas@iitk.ac.in, bidyutd@iitk.ac.in, surajprk@iitk.ac.in, shakel@iitk.ac.in}

\affiliation{ Indian Institute of Technology - Kanpur, Kanpur 208016, Uttar Pradesh, India}

\abstract{We work perturbatively with an interacting quantum field theory comprised of two distinct scalar fields. In this theory, we introduce a sudden quench of the mass of one of the scalars at time $t_0$. Also, the quartic interaction between the two scalars is turned on at time $t_{in}$. These break time-translation invariance. In this setup we examine the effects of the relative ordering of $t_0$ and $t_{in}$ on composite operator mixing. We study how such operator mixing affect features of the scalar potential. We find that the late time effective potential can be sensitive enough to the quenches to trigger phase transitions.}

\begin{document}
\setstretch{1.2}
\maketitle

\section{Introduction}

We exist \textit{out of equilibrium}, yet most physics tools are mostly developed keeping in mind equilibrium scenarios. Our universe is expanding and hence the quantum fluctuations on top of it are essentially describable via time-dependent couplings which drive the theory into non-equilibrium regimes. This is the regime of quantum quenches. The effective field theory is  described by the finite set of operators which are commensurate with the symmetries, and in equilibrium, one has the luxury of Poincare invariance: which limits operators to local ones. However, when the high energy theory becomes subjected to non-equilibrium (as with time-dependent couplings arising in an expanding spacetime) then one does not have time translation symmetry, and the effective theory is no longer local. 

 The quantum quench, which specifies the time dependence of the couplings can either be sudden or smooth with a characteristic time scale. The physics of quenches have recently garnered a lot of attention in condensed matter systems due to controllable cold atom experiments. In these set-ups critical quenches have been carried out which result in universal scalings as predicted by Kibble and Zurek \cite{Kibble:1976sj, Zurek:1985qw}. There are very few analytically tractable examples for KZ scalings and hence most studies have been limited to either free field theories \cite{Das:2014hqa, Das:2014jna, Das:2015jka, Das:2016lla, Das:2017sgp} or exactly solvable large N models \cite{Das2012, Das:2020dfe}, or holographic set-ups \cite{Das:2016eao}. The mechanism for these scalings are still poorly understood since they focus on observables which are strongly theory dependent. The observables for time-dependent systems are correlation functions instead of scattering amplitudes, since unlike in a time-translation invariant system the S-matrix is not definable.

In this work, we look at the structure of composite operators in the non-equilibrium theory and try to draw universal lessons. In particular, we consider a generically interacting theory whose couplings are quenched independently. This essentially excites the system, taking us away from adiabaticity. Therefore, in this setting even if we want to integrate out certain heavy field loops, we cannot completely get rid of them from the effective low energy theory. This is because the excitement caused due to the quench also creates heavy field excitations, which are part of the {\em in-}state \cite{Dymarsky:2017awt}. Hence, the time-evolving light field operators, which are now computed using the Schwinger-Keldysh contour or the {\em in-in} formalism, involve the creation and annihilation of heavy fields. There is a non-trivial operator mixing which contains detailed information about the quench.  There are new types of divergences involving time derivatives of the quenched couplings, and RG flows due to a particular quench may trigger quench protocols of other couplings at different scales \cite{Goykhman:2018iaz}. Renormalization in the presence of time-dependent couplings also arises in the general case of QFT in curved spacetime as explored in \cite{Brown:1980qq, Birrell:1982ix}.

Such operator mixings have previously also been explored in the context of the interacting double scalar ($g^2 \phi^2 \chi^2$) model \cite{Dresti:2013kya}. The interaction gets turned on at a particular time $t_{in}$, which is how the system is driven into non-equilibrium. The authors found by looking into {\em in-in} correlators that the $\phi^2(t,x)$ operator (at leading order in interaction coupling) mixes with $\chi^2(t)$ as well as with $\chi^2(t_{in})$. However, the renormalization group conditions can be chosen in a way such that the $\chi^2(t)$ mixing vanishes at the one loop order, though there is no way to get rid of $\chi^2(t_{in})$ from the mixing. This latter mixing is a signature of{ \em memory} of the quench event. The coefficient of this term, the kernel $K(t-t_{in})$, is explicitly non-local, and shows a power law decay in $m_\phi (t-t_{in})$, where $m_\phi$ is the mass of the $\phi$ field.  At very late-times, with proper RG conditions, there is therefore no non-trivial mixing (at ${\cal O}(g^2)$), and hence no tell-tale signatures of the quench in the relevant effective potential.
In the setup of our present work, we work with the same double scalar model, however, making departures in two different directions:
\begin{itemize}
	\item We introduce a mass quench in one of the fields. This is done independently from the interaction quench, viz. the mass quench occurs at some time $t_0$ which is generically different from the time of the interaction quench.
	\item The effect of the mass quench is treated non-perturbatively. This is both important (as the mass operator is more relevant) as well as possible (as the time-dependent free theory is still quadratic ). 
\end{itemize}

Our results show a variety of interesting features, some of which we summarize below:
\begin{enumerate}
	\item There is memory of both the quench events: the mass quench as well as the interaction quench in the operator mixing. In the composite operator $\phi^2(t,\vec{x})$ the signature of this is the presence of $\chi^2(t_{in})$ which comes with its own non-local memory kernels. The kernels are different depending on the order of the quenches. However, both these kernels decay at large times away from the quench events and hence disappear from late-time perturbative physics. The presence of memory right after the quench is a signature of the deep non-Markovian characteristic of non-equilibrium QFT \cite{Burrage:2018pyg}.
	
	\item As with the interaction quench, we once again have the mixing with the local operator: $\chi^2(t)$. However, unlike \cite{Dresti:2013kya}, this time there is no RG condition which can make this go away. This is because, in addition to the logarithmic divergence, the coefficient also contains finite terms arising purely due to the mass quench. This term, therefore, affects late-time physics. Thus, the signatures get carried over into the effective potential, and hence these very early-time quenches can play a crucial role in deciding the late-time phase structure of the quantum field configuration. This possibility is very tantalizing when put in the context of quantum fluctuations during inflation.
\end{enumerate}

In the sections to follow, we proceed with the following outline: In \S \ref{sec:time-dep-SHO} we work out the non-equilibrium Greens function for a time-dependent quantum harmonic oscillator. This will directly apply to the scalar field theory with the mass quench and will allow us to determine the free propagators along the Keldysh contours. \S \ref{sec:operator-mixing} introduces the model that we focus on and the quench protocols. After defining the Feynman rules on the Keldysh contour, we proceed in this section to compute the one loop effect of the interaction on operator mixings for the different quench sequences. Next in \S \ref{sec:analysis} we discuss the renormalization and late-time analytical form of the perturbative mixing. We also evaluate various contributions to the composite operator numerically and point out crucial dependencies on the details of the quench protocol. \S \ref{sec:chi-potential} deals with the impact of the non-trivial mixing on the effective potential and explicitly analyzes how the mass quench can affect the late-time phase diagram of the theory. We end with conclusions in \S \ref{sec:conclusion}. 

\section{Simple harmonic oscillator with time dependent Hamiltonian}\label{sec:time-dep-SHO} 

In momentum space, the free scalar field theory is a set of independent quantum harmonic oscillators. We will need the non-equilibrium Green's functions for the free scalar field due to mass quench. Hence, in this section, we compute the analogous correlators for the single oscillator, which we will employ to compute the Green's functions for the free scalar field. If the frequency of a simple harmonic oscillator is suddenly quenched from $\omega_0$ to $\omega$ at time $t = t_0$ \cite{Cardy:2010si,BenTov:2021jsf}, then its Hamiltonian carries an explicit time-dependence which can be described as
\begin{eqnarray}
H(t) = \frac{1}{2} p^2 + \frac{\omega^2(t)}{2} q^2,
\end{eqnarray}
with $\omega^2(t) = \theta(t_0 - t)\,\omega_0^2 + \theta(t- t_0)\, \omega^2$. The time dependence is captured within the Heaviside theta functions. Away from $t_0$, the Hamiltonian in each region (for $t < t_0$ as well as for $t > t_0$) assumes the usual time-independent form, but with different frequencies ($\omega_0$ for $t < t_0$ and $\omega$ for $t > t_0$) in the two regions. Therefore, the initial Hamiltonian before quench $(t<t_0)$ is
\begin{eqnarray*}
	H_{in}&=&\frac{1}{2}p^2+ \frac{1}{2}\,\omega_0^2\, x^2.
\end{eqnarray*}

In the Heisenberg picture, the time evolution of position operator $x$ for $t<t_0$ can be expressed in terms of creation and annihilation operators, $a^\dagger_{in}$ and $a_{in}$, as
\begin{equation}\label{eq:q-before}
	x(t<t_0)=\frac{1}{\sqrt{2\omega_0}}\left(e^{-i\omega_0(t-t_0)}a_{in}+e^{i\omega_0(t-t_0)}a^\dagger_{in}\right),
\end{equation}
where the annihilation operator acts on the ground state configuration of the initial states to give $0$, i.e., $a_{in} | 0_{in} \rangle = 0$. This allows us to recast the Hamiltonian as 
\begin{eqnarray}
H_{in} = (a^\dagger_{in}a_{in}+\frac{1}{2})\, \omega_0.
\end{eqnarray}
\noindent
The expression for the final Hamiltonian, after the quench $(t>t_0)$, is given as
\begin{eqnarray*}
	H_{out}&=&\frac{1}{2}p^2+ \frac{1}{2}\omega^2 x^2.
\end{eqnarray*}
Once again, the Heisenberg picture position operator in $t > t_0$ can again be expressed as
\begin{equation}\label{eq:q-after}
	x(t>t_0)=\frac{1}{\sqrt{2\omega}}\left(e^{-i\omega(t-t_0)}a_{out}+e^{i\omega(t-t_0)}a^\dagger_{out}\right),
\end{equation}
where $a_{out} \,| 0_{out} \rangle = 0$ since $|0_{out} \rangle$ is the ground state of the final Hamiltonian. It is obvious that $ |0_{in} \rangle \neq |0_{out}\rangle$. $H_{out}$ can be rewritten in terms of $a^\dagger_{out}$ and $a_{out}$ as
\begin{eqnarray}
H_{out} = (a^\dagger_{out}a_{out}+\frac{1}{2})\,\omega.
\end{eqnarray}

 \noindent When $\omega_0 = \omega$ then $a_{in} = a_{out}$, otherwise they are different. However using a Bogoliubov transformation one can recast $a_{out}$ as a linear combination of $a_{in}$ and $a^\dagger_{in}$ as shown below:
\begin{eqnarray}\label{eq:aout-ain}
	a_{out}=A_1 \, a_{in} + A_2 \, a^\dagger_{in}.
\end{eqnarray}
We compute the Bogoliubov coefficients by matching the Heisenberg operators at quench time $t_0$. Explicit computation of the Bogoliubov coefficients yields \cite{BenTov:2021jsf}:
\begin{equation}\label{eq:A12value}
	A_1\,=\, \frac{\omega-\omega_0}{2\sqrt{\omega\,\omega_0}}, \quad\quad\quad 	A_2\,=\, \frac{\omega+\omega_0}{2\sqrt{\omega\,\omega_0}}.
\end{equation}
Using the above result to replace $A_{1,2}$ by coefficients dependent on $\omega_0$ and $\omega$ in Eq.~\eqref{eq:aout-ain} and substituting for $a_{out}$ in Eq.~\eqref{eq:q-after} we get,
\begin{equation}\label{eq:q-after-final}
	x(t>t_0)=\left[u_{in}(t)\,a_{in}\,+\,u_{in}^{*}(t)\,a^\dagger_{in}\right],
\end{equation}
with,
\begin{equation}\label{eq:u-in-t}
	u_{in}(t)=\frac{1}{\sqrt{2\omega_0}}\left[\cos (\omega(t-t_0)) -\frac{i\omega_0}{\omega}\sin(\omega(t-t_0))\right].
\end{equation}

\noindent Combining the position operators, see Eqs.~\eqref{eq:q-before} and \eqref{eq:q-after-final} for the two regions ($t < t_0$ and $t > t_0$) using Heaviside $\theta$-functions we obtain,
\begin{eqnarray}\label{eq:q-final}
	x(t)= & & \left(\theta(t_0-t)\frac{e^{-i\omega_0(t-t_0)}}{\sqrt{2\omega_0}}+\theta(t-t_0)u_{in}(t)\right)a_{in} \nonumber \\ 
	& &\,+ \left(\theta(t_0-t)\frac{e^{i\omega_0(t-t_0)}}{\sqrt{2\omega_0}}+\theta(t-t_0)u_{in}^{*}(t)\right)a^\dagger_{in}. 
\end{eqnarray}
The expectation value of two position operators at different times $t_1$ and $t_2$  in the in-vacuum state takes the form as
\begin{eqnarray}\label{eq:gpm}
	\langle 0_{in}| x(t_1) x(t_2) |0_{in}\rangle = & & \left(\theta(t_0-t_1)\frac{e^{-i\omega_0(t_1-t_0)}}{\sqrt{2\omega_0}}+\theta(t_1-t_0)u_{in}(t_1)\right) \nonumber \\ 
	& &\,\times \left(\theta(t_0-t_2)\frac{e^{i\omega_0(t_2-t_0)}}{\sqrt{2\omega_0}}+\theta(t_2-t_0)u_{in}^{*}(t_2)\right).
\end{eqnarray}
It is worth mentioning that always in the limit $ \omega=\omega_0 $, i.e. for no quench, the above two-point correlation function assumes the same form as the correlation function of the two position operators at different times for a harmonic oscillator with time-independent frequency, i.e.,
\begin{eqnarray}\label{key}
	\langle 0_{in}| x(t_1) x(t_2) |0_{in}\rangle =\frac{1}{2\,\omega}e^{-i\omega(t_1-t_2)}.
\end{eqnarray}

\section{Composite operator mixing through (non-)local kernels}\label{sec:operator-mixing}

As alluded to in the introduction, the operator correlators are the observables in the {\it in-in } formalism. In a quantum field theory, there are infinite operators which can be organized in scales of relevance. Among the members of this infinite set, the composite operator built by squaring the fundamental field: $\phi^2(t,\vec{x})$, is the simplest non-trivial one, which  contributes to the energy. In an interacting QFT this operator generically exhibits non-trivial mixing with other fundamental fields in the theory. This arises when, in the connected Feynman diagrams involving $\phi^2$, there are loops consisting of the other fundamental fields. Clearly this originates due to interaction between the different fundamental fields, and usually results in non-trivial renormalization group flow of the composite operator. The flow decides among other things the measurable critical exponents associated with various physical observables where $\phi^2$ contributes. It is far from understood how operator mixing takes place in \textit{out of equilibrium}. 
In what follows, we have studied the mixing of $\phi^2$ in systems where time-translation invariance is broken explicitly by multiple quantum quenches. 

\subsection{Model description and Green's functions}

We consider a simplified framework consists of two real scalar fields $\phi$ and $\chi$ with different mass parameters, and the  Lagrangian for the system can be written as
\begin{eqnarray}\label{eq:lag}
	\mathcal{L} &=& \mathcal{L}_0 [\phi, \chi] + \mathcal{L}_{\text{int}}[\phi,\chi] - \frac{\lambda_\phi}{4 !} \phi^4 - \frac{\lambda_\chi}{4 !} \chi^4,
\end{eqnarray}
where $\mathcal{L}_0$ describes the free-field Lagrangian and $\mathcal{L}_{\text{int}}$ encapsulates the interaction between the two fields, i.e.,
\begin{eqnarray}
	\mathcal{L}_0 [\phi, \chi] &=& \frac{1}{2}(\partial_\mu\phi \partial^\mu\phi) -\frac{1}{2}m^2(t)\,\phi^2 +\frac{1}{2}(\partial_\mu\chi \partial^\mu\chi) -\frac{1}{2}M^2\chi^2, \nonumber \\
	\mathcal{L}_{\text{int}}[\phi,\chi] &=& -\frac{g^2}{2}\phi^2\chi^2 .
\end{eqnarray} 
The only internal symmetry respected by the Lagrangian is a $\mathbb{Z}_2$ symmetry for each of the fields that filters out terms containing odd powers of either field from the Lagrangian. Terms such as quartic self-interaction of the fields does not contribute in the operator mixing of $\phi^2$ but will be of significance when the impact of composite operator mixing on the scalar potential is discussed. Also, the term linear in both fields, i.e., the $\phi\,\chi$ coupling can be removed through global rotations of the two fields followed by field redefinitions.
\\
\\
\noindent The breaking of time-translation invariance has been accomplished in two ways:  
\begin{enumerate}
\item  By imposing an explicit initial time $t_{in}$ as the lower limit of the time integral present in the path integral. Since we are looking at the composite operator $\phi^2$, this is equivalent to quenching the interaction term by suddenly turning on the interaction at time $t = t_{in}$ \cite{Dresti:2013kya}. Based on this, one can write the action as,
\begin{eqnarray}
S[\phi, \chi] = \int_{-\infty}^{\infty} dt \int d^3 x \,\big( \mathcal{L}_0 [\phi, \chi]  + \Theta(t - t_{in})  \, \mathcal{L}_{\text{int}} [\phi, \chi] \big).
\end{eqnarray}

\item Additionally, we suddenly change the mass of the $\phi$ field from $m_0$ to $m$ at a time $t = t_0$ which we have referred to as the mass quench of the scalar field throughout the paper.
\end{enumerate}

\begin{figure}
	\centering
	\includegraphics[width=0.7\linewidth]{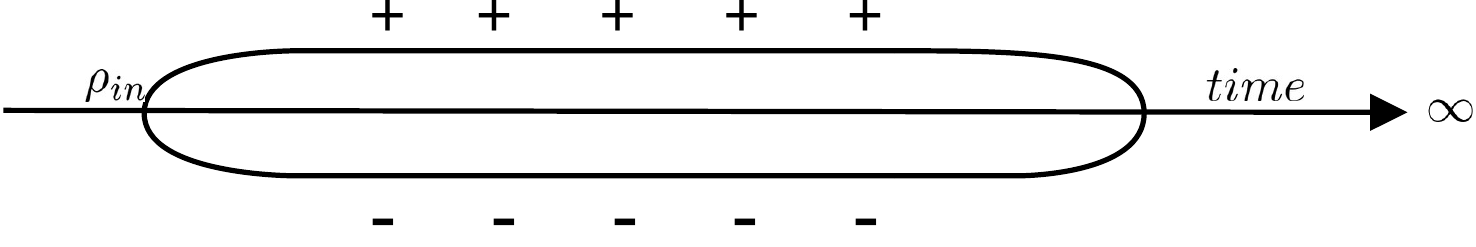}
	\caption{Closed time contour $ \mathcal{C} $.}
	\label{fig:skcontour}
\end{figure}

\noindent The breaking of time-translation invariance prompts us to follow the Schwinger-Keldysh (or the \emph{in-in}) formulation. This results in doubling the degrees of freedom for both $\phi(x)$ and $\chi(x)$ (through the introduction of $\phi_{\pm}(x)$ and $\chi_{\pm}(x)$). On the Schwinger-Keldysh contour \ref{fig:skcontour}, the action is expressed as:
\begin{eqnarray}
	S & = & \int_{t_{in}}^{\infty} dt \int d^3x \Big[\mathcal{L}[\phi_+,\chi_+]-\mathcal{L}[\phi_-,\chi_-]\Big] \nonumber\\
	& = & \int_{t_{in}}^{\infty} dt \int d^3x \Big[ \frac{1}{2}(\partial_\mu\phi_+)(\partial^\mu\phi_+)  -\frac{1}{2}m^2\phi_+^2 +\frac{1}{2}(\partial_\mu\chi_+)(\partial^\mu\chi_+) -\frac{1}{2}M^2\chi_+^2 -\frac{g^2}{2}\phi_+^2\chi_+^2 \nonumber\\
	& &\qquad\quad -\frac{1}{2}(\partial_\mu\phi_- )(\partial^\mu\phi_-) +\frac{1}{2}m^2\phi_-^2 -\frac{1}{2}(\partial_\mu\chi_-)( \partial^\mu\chi_-) +\frac{1}{2}M^2\chi_-^2+\frac{g^2}{2}\phi_-^2\chi_-^2  \Big]. 
\end{eqnarray}

\noindent
Free scalar field theory can be described as the superposition of independent momentum modes, each of which evolves as a simple harmonic oscillator. Consequently, the propagators of a scalar field and that of a harmonic oscillator can be related by a Fourier transform, 
\begin{eqnarray}\label{eq:gmp-gpm-phi}
	G_{-+}^{\phi}(x , y) &=& \langle \phi(x) \phi(y)\rangle  = \int \frac{d^3\vec{k} }{(2\pi)^3} e^{-i \vec{k}\cdot (\vec{x} - \vec{y})}\, G_{- +}^{\phi}(\vec{k}, t_1, t_2), \nonumber\\
	G_{+-}^{\phi}(x , y) &=& \langle \phi(y) \phi(x)\rangle  = \int \frac{d^3\vec{k} }{(2\pi)^3} e^{-i \vec{k}\cdot (\vec{x} - \vec{y})}\, G_{+-}^{\phi}(\vec{k}, t_1, t_2).
\end{eqnarray}
\noindent
Here, $x = (t_1, \vec{x})$, $y = (t_2, \vec{y})$, $t_1 \equiv x^0$ and $t_2 \equiv y^0$. We assume that the initial density matrix $\rho_{in}$ is the vacuum state of free field theory. The Green's functions in the momentum space are expressed as (see Eq.~\eqref{eq:gpm})
\begin{eqnarray}\label{eq:gpmk}
	G_{-+}^{\phi}(\vec{k}, t_1,t_2)&=& \left(\theta(t_0-t_1)\frac{e^{-i\omega_{0k}(t_1-t_0)}}{\sqrt{2\omega_{0k}}}+\theta(t_1-t_0)u_{in}(\vec{k}, t_1)\right) \times \nonumber \\ 
	&& \left(\theta(t_0-t_2)\frac{e^{i\omega_{0k}(t_2-t_0)}}{\sqrt{2\omega_{0k}}}+\theta(t_2-t_0)u_{in}^{*}(\vec{k},t_2)\right) =  G_{+-}^{\phi}(\vec{k}, t_2,t_1).
\end{eqnarray}
\noindent
In the above equation, $u_{in}(\vec{k},t_i)$ and $u_{in}^{*}(\vec{k},t_i)$, $i=1,2$ are the functions defined in Eq.\eqref{eq:u-in-t} with the momentum dependence explicitly highlighted, i.e.,
\begin{eqnarray}\label{eq:u-in-kt}
	u_{in}(\vec{k}, t)=\frac{1}{\sqrt{2\omega_{0 k}}}\left[\cos (\omega_k(t-t_0)) -\frac{i\omega_{0k}}{\omega_k}\sin(\omega_k(t-t_0))\right].
\end{eqnarray}

Here, $\omega_{0 k} = \sqrt{\vec{k}^2  + m_0^2}$ and $\omega_k = \sqrt{\vec{k}^2+m^2}$ are the frequencies of the $\phi$ field before and after the mass quench. The propagators for $\chi$ fields can similarly be written as: 
\begin{eqnarray}\label{eq:gmp-gpm-chi}
	G_{-+}^{\chi}(x , y) &=& \langle \chi(x) \chi(y)\rangle = \int \frac{d^3\vec{k} }{(2\pi)^3} e^{-i \vec{k}\cdot (\vec{x} - \vec{y})} \, G_{- +}^{\chi}(\vec{k}, t_1, t_2), \nonumber\\
	G_{+ -}^{\chi}(x , y) &=& \langle \chi(y) \chi(x)\rangle = \int \frac{d^3\vec{k} }{(2\pi)^3} e^{-i \vec{k}\cdot (\vec{x} - \vec{y})} \, G_{+ -}^{\chi}(\vec{k}, t_1, t_2).
\end{eqnarray}
In this case, the momentum space Green's functions for the $\chi$ fields have the following simple form:
\begin{eqnarray}\label{eq:gmp-gpm-chi-2}
	G_{- +}^{\chi}(\vec{k},t_1, t_2) = \frac{e^{-i\Omega_{k}(t_1 - t_2)}}{2\Omega_{k}} = G_{+ -}^{\chi}(\vec{k},t_2, t_1),
\end{eqnarray}
\noindent
with $\Omega_{k} = \sqrt{\vec{k}^2 + M^2}$ being the frequency. The time ordered and anti-time ordered propagators of the fields can be expressed respectively by the linear combinations of the Green's functions defined in Eqs.~\eqref{eq:gmp-gpm-phi} and \eqref{eq:gmp-gpm-chi} as
\begin{eqnarray}\label{eq:gpp-gmm}
	G^{\phi,\,\chi}_{++}(x,y) &=& \theta(x^0 - y^0)\, G^{\phi,\,\chi}_{-+}(x,y) + \theta(y^0 - x^0)\, G^{\phi,\,\chi}_{+-}(x,y), \nonumber\\
	G^{\phi,\,\chi}_{--}(x,y) &=& \theta(x^0 - y^0)\, G^{\phi,\,\chi}_{+-}(x,y) + \theta(y^0 - x^0)\, G^{\phi,\,\chi}_{-+}(x,y).
\end{eqnarray}

\begin{figure}[!htb]
	\centering
	\renewcommand\thesubfigure{\roman{subfigure}}
	\begin{subfigure}[t]{3cm}
		\centering
		\includegraphics[trim= 0cm 1.5cm 0cm 0cm ,height=2cm, width=3.4cm]{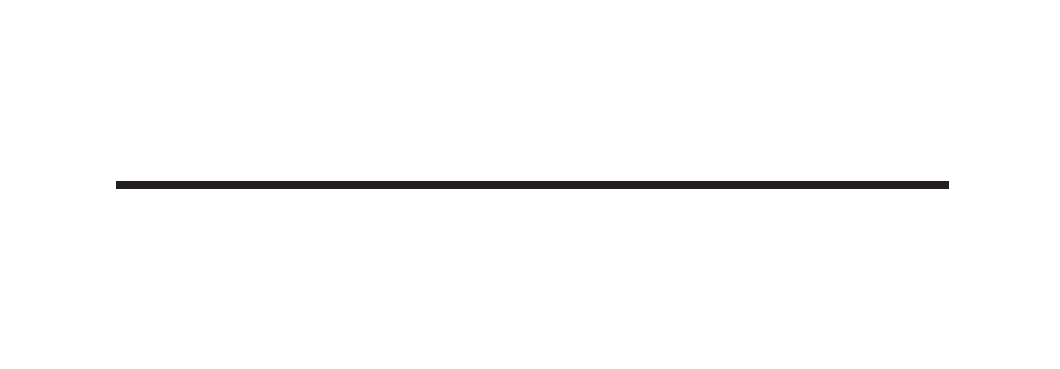}
		\caption{$\phi_+$}
	\end{subfigure}\hspace{0.5cm}
	\begin{subfigure}[t]{3cm}
		\centering
		\includegraphics[trim= 0cm 1.5cm 0cm 0cm, height=2cm, width=3.4cm]{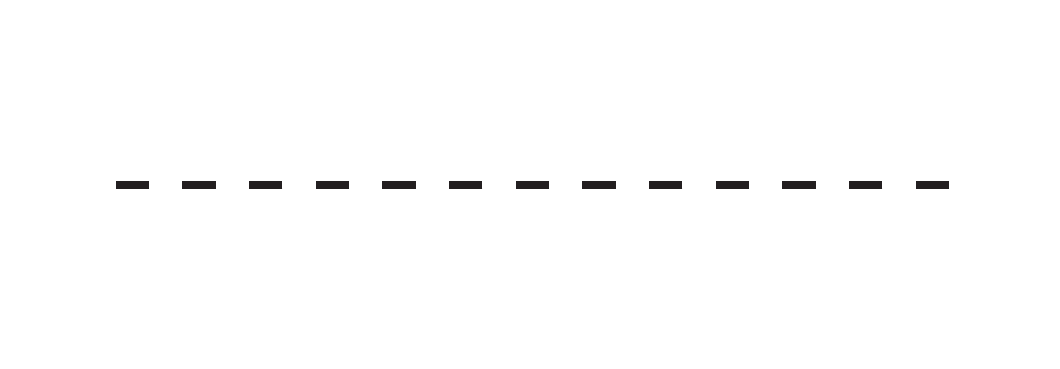}
		\caption{$\phi_-$}
	\end{subfigure}\hspace{0.5cm}
	\begin{subfigure}[t]{3cm}
		\centering
		\includegraphics[trim= 0cm 1.5cm 0cm 0cm ,height=2cm, width=3.4cm]{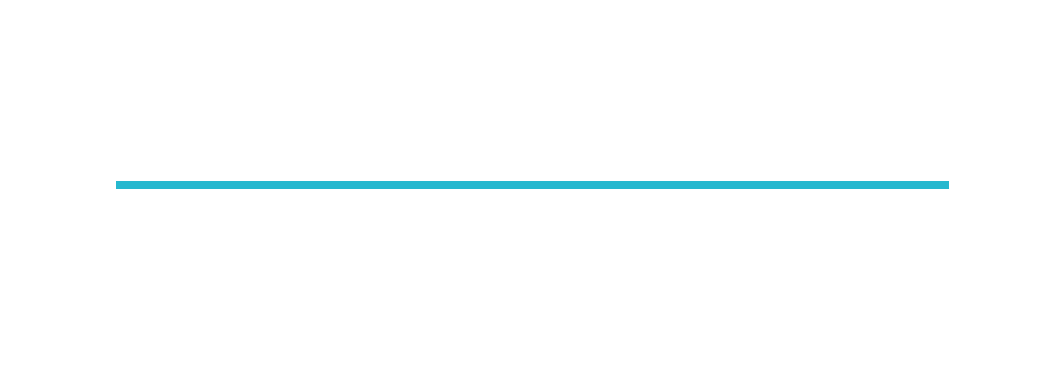}
		\caption{$\chi_+$}
	\end{subfigure}\hspace{0.5cm}
	\begin{subfigure}[t]{3cm}
		\centering
		\includegraphics[trim= 0cm 1.5cm 0cm 0cm, height=2cm, width=3.4cm]{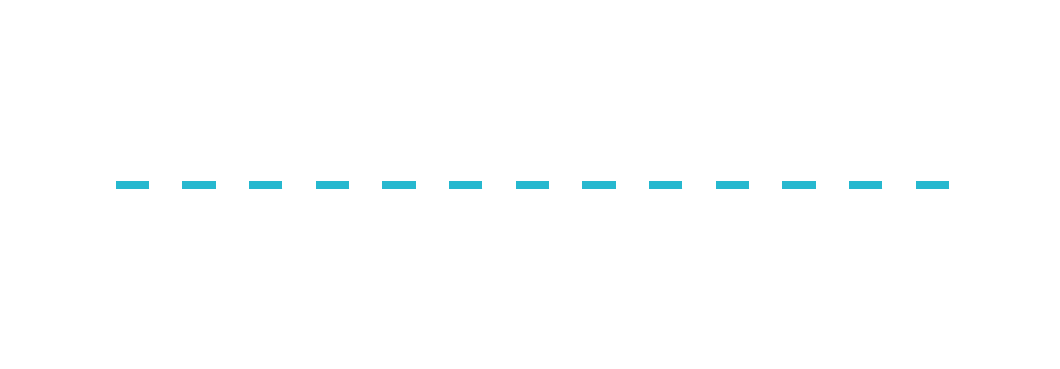}
		\caption{$\chi_-$}
	\end{subfigure}
	\caption{\sf Symbols corresponding to the $\phi_\pm$, and $\chi_\pm$ fields. }
	\label{fig:propagators}
\end{figure}

\begin{figure}[!htb]
	\centering
	\renewcommand\thesubfigure{\roman{subfigure}}
	\begin{subfigure}[t]{7cm}
		\centering
		\includegraphics[trim= 0cm 1.5cm 0cm 0cm ,height=1.6cm, width=6.4cm]{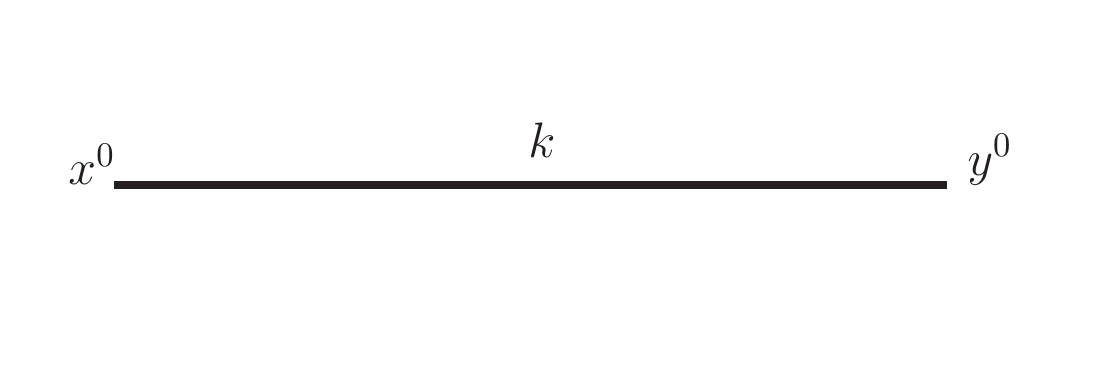}
		\caption{$G^\phi_{++}(\vec{k},x^0,y^0)$}
	\end{subfigure}
	\begin{subfigure}[t]{7cm}
		\centering
		\includegraphics[trim= 0cm 1.5cm 0cm 0cm, height=1.6cm, width=6.4cm]{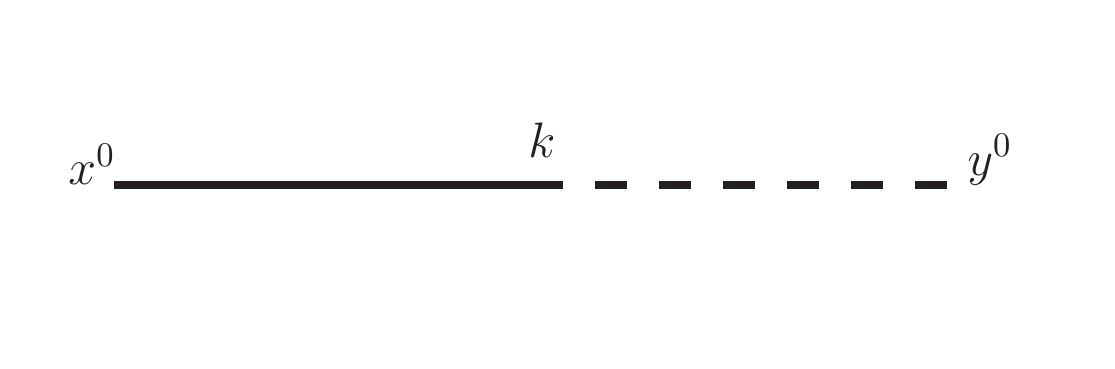}
		\caption{$G^\phi_{+-}(\vec{k},x^0,y^0)$}
	\end{subfigure}
	\noindent\begin{subfigure}[t]{7cm}
		\centering
		\includegraphics[trim= 0cm 1.5cm 0cm 0cm ,height=1.6cm, width=6.4cm]{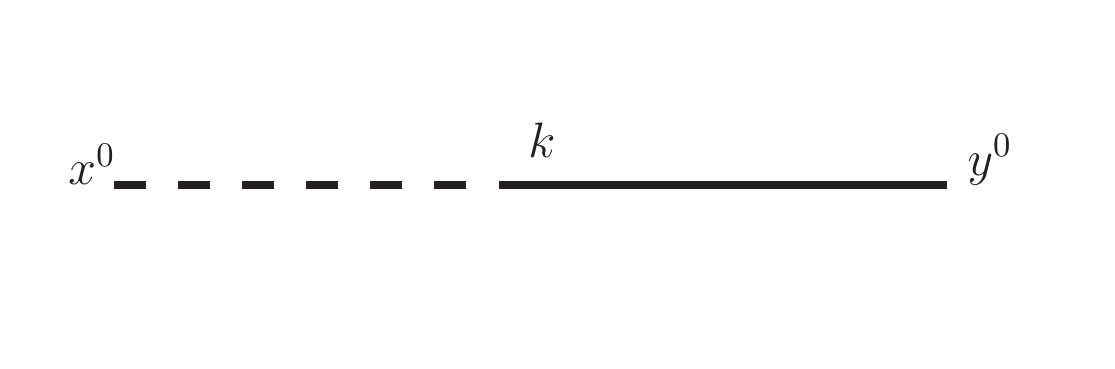}
		\caption{$G^\phi_{-+}(\vec{k},x^0,y^0)$}
	\end{subfigure}
	\begin{subfigure}[t]{7cm}
		\centering
		\includegraphics[trim= 0cm 1.5cm 0cm 0cm, height=1.6cm, width=6.4cm]{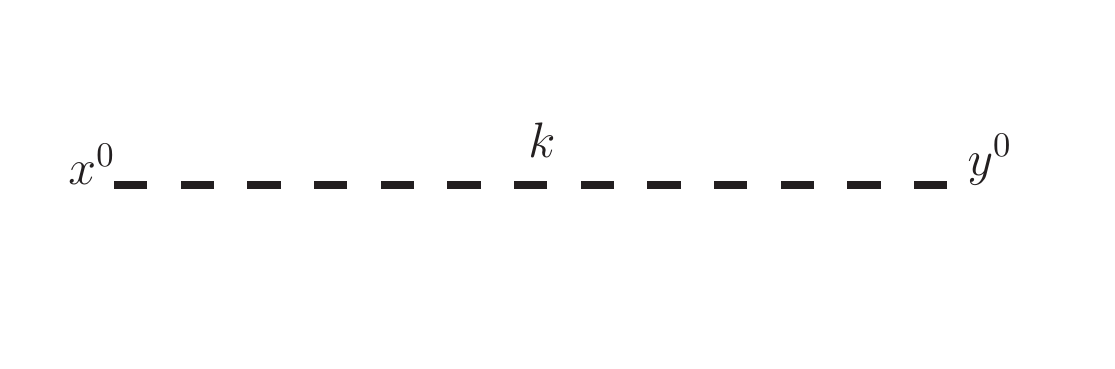}
		\caption{$G^\phi_{--}(\vec{k},x^0,y^0)$}
	\end{subfigure}
	\noindent\begin{subfigure}[t]{7cm}
		\centering
		\includegraphics[trim= 0cm 1.5cm 0cm 0cm ,height=1.6cm, width=6.4cm]{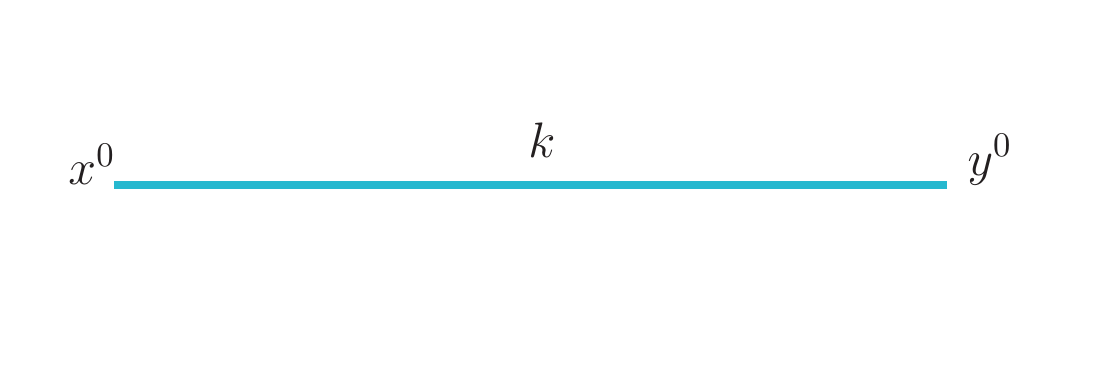}
		\caption{$G^\chi_{++}(\vec{k},x^0,y^0)$}
	\end{subfigure}
	\begin{subfigure}[t]{7cm}
		\centering
		\includegraphics[trim= 0cm 1.5cm 0cm 0cm, height=1.6cm, width=6.4cm]{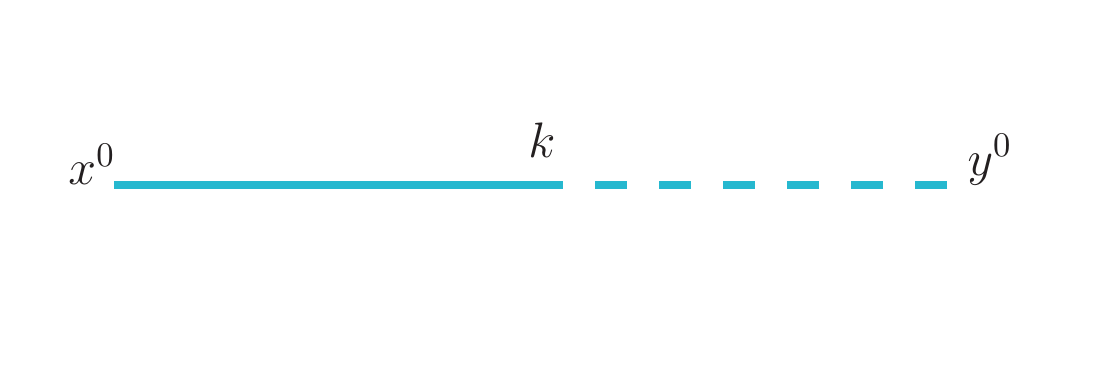}
		\caption{$G^\chi_{+-}(\vec{k},x^0,y^0)$}
	\end{subfigure}
	\noindent\begin{subfigure}[t]{7cm}
		\centering
		\includegraphics[trim= 0cm 1.5cm 0cm 0cm ,height=1.6cm, width=6.4cm]{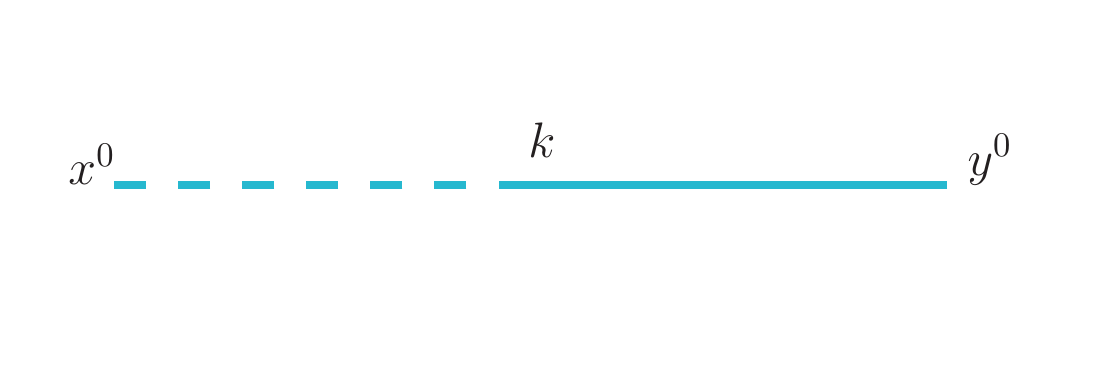}
		\caption{$G^\chi_{-+}(\vec{k},x^0,y^0)$}
	\end{subfigure}
	\begin{subfigure}[t]{7cm}
		\centering
		\includegraphics[trim= 0cm 1.5cm 0cm 0cm, height=1.6cm, width=6.4cm]{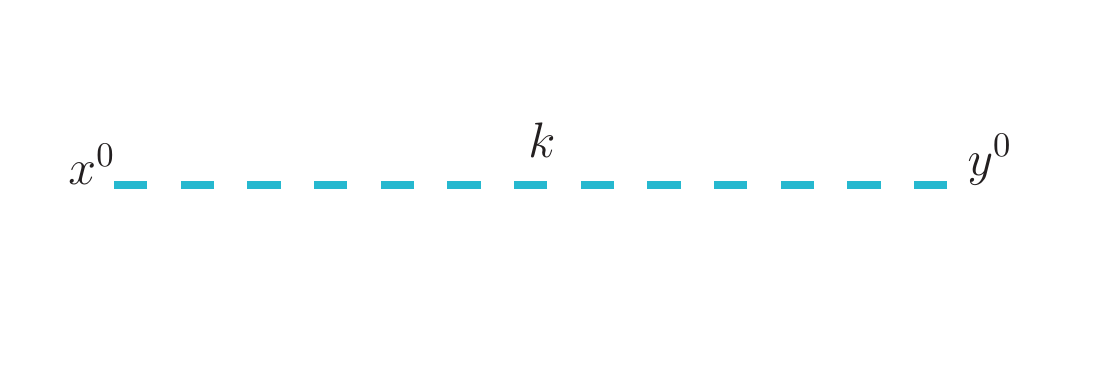}
		\caption{$G^\chi_{--}(\vec{k},x^0,y^0)$}
	\end{subfigure}
	\caption{\sf Diagrammatic representation of Green's functions $G^{\phi,\,\chi}_{ij}(x^0, y^0)$ with $i,\,j \in  \{+, -\}$.}
	\label{fig:greens-functions}
\end{figure}

\begin{figure}[!htb]
	\centering
	\renewcommand\thesubfigure{\roman{subfigure}}
	\begin{subfigure}[t]{5cm}
		\centering
		\includegraphics[scale=0.48]{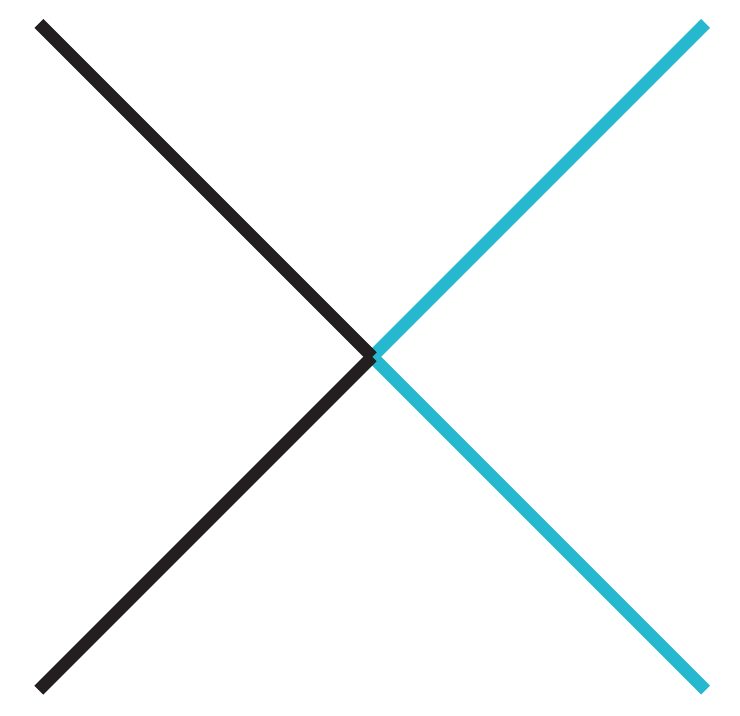}
		\caption{$\phi^2_+\chi^2_+,\,\,\,$   vertex factor: $-\cfrac{i g^2}{2}$}
	\end{subfigure}\hspace{1cm}
	\begin{subfigure}[t]{5cm}
		\centering
		\includegraphics[scale=0.48]{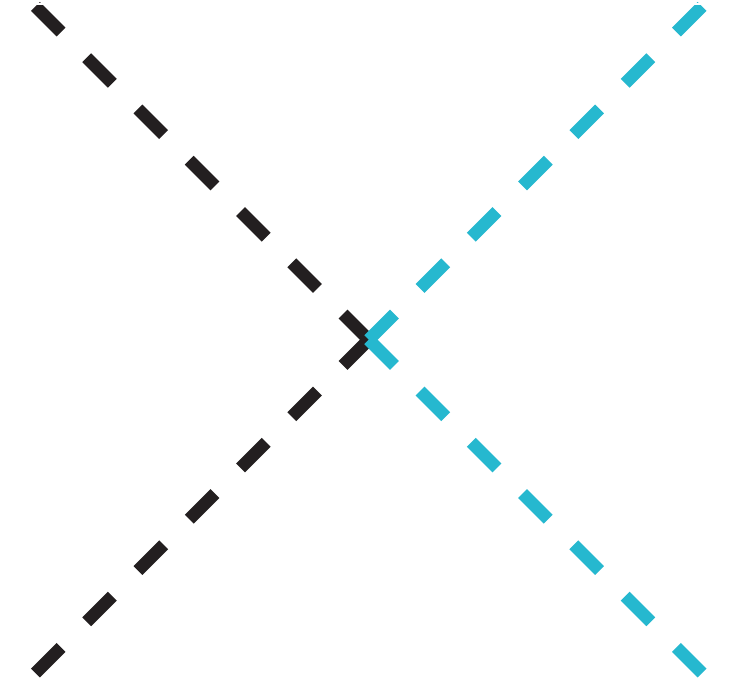}
		\caption{$\phi^2_-\chi^2_-,\,\,\,$   vertex factor: $\cfrac{i g^2}{2}$}
	\end{subfigure}
	\caption{\sf Feynman diagrams corresponding to contact interactions. }
	\label{fig:vertices}
\end{figure}

The Feynman rules corresponding to the $\phi_{\pm}$ and $\chi_{\pm}$ fields, the different propagators of $\phi$ and $\chi$, $G^{\phi,\,\chi}_{ij}$ fields with $i,\,j \in  \{+, -\}$, and for the vertices of the quartic interactions - $\phi^2_+\,\chi^2_+$ and $\phi^2_+\,\chi^2_-$ have been depicted in Figs.~\ref{fig:propagators}, \ref{fig:greens-functions} and \ref{fig:vertices} respectively.

\subsection{Effect of the chronology of quenches on operator mixing}

In \cite{Dresti:2013kya}, an expression for $\phi^2(t)$ in terms of $\chi^2(t)$, $\chi^2(t_0)$, $\chi^2(t_{in})$ was obtained when time-translation symmetry was broken by explicitly switching on the $\phi$ -$\chi$ quartic interaction at time $t_{in}$.

Here, we outline the effect of introducing a sudden quench of the mass of $\phi$ field ($m_0 \rightarrow m$) at $t_0$ in addition to switching on the quartic interaction at $t_{in}$ and highlight the differences between two specific cases  - (1.) $t_0 < t_{in}$ and (2.) $t_{in} < t_0$. 

To obtain the expression for $\phi^2(t)$ in terms of $\chi^2(t)$, $\chi^2(t_0)$, $\chi^2(t_{in})$, it is necessary to first compute the sum of  all possible connected Green's functions of the form $\langle \phi_i (t)\phi_j(t)\chi_k(t_1)\chi_l(t_2)\rangle_c$ (the subscript c denotes connected correlators) with $i, j, k, l \in \{+, - \}$. Thus there will be $2^4 = 16$ different correlation functions. It must be noted that in both the cases taken into account in our analysis: $t_0 < t_{in}$ or $t_{in} < t_0$, the other time instances involved in the four-point correlation function $\langle \phi_i (t)\phi_j(t)\chi_k(t_1)\chi_l(t_2)\rangle_c$, i.e., $t,\,t_1,\,t_2$ always maintain the chronology $t_0, t_{in}\, < \, t \, <\, t_1\,<\,t_2$. We restrict ourselves to the cases with $t > t_0, t_{in}$ because we are interested in late-time physics. Also, $t_1, t_2 > t$ must hold so that there is no effect of the external states on the operator mixing. 

\begin{figure}[!htb]
	\centering
	\renewcommand\thesubfigure{\roman{subfigure}}
	\includegraphics[scale=0.75]{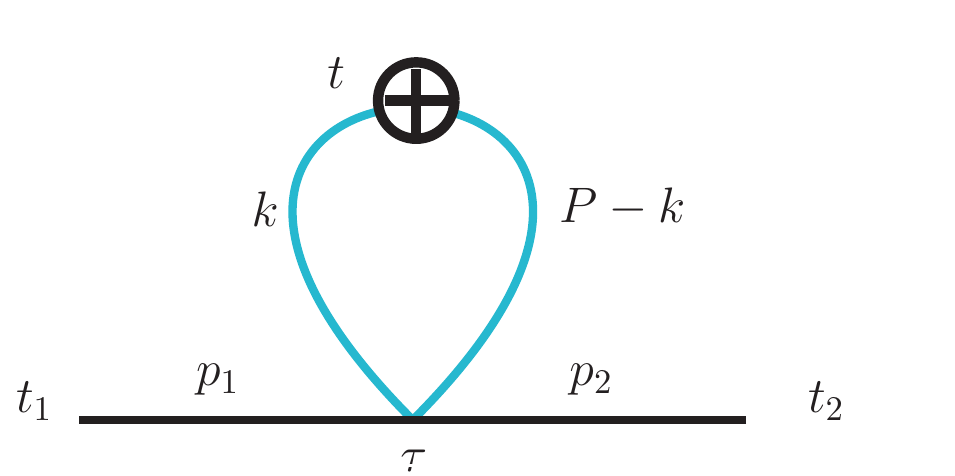} 
	\hspace{0.1cm}\includegraphics[scale=0.75]{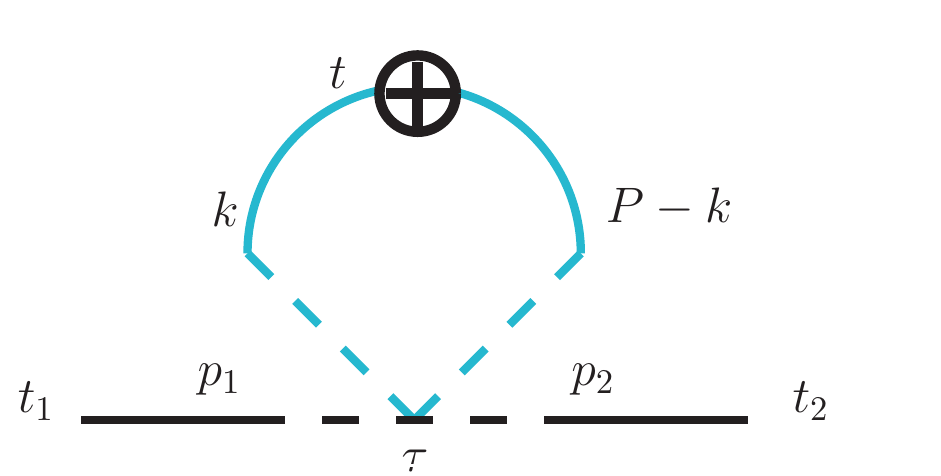} 
	\caption{\sf One-loop diagrams (i) $L_1$ and (ii) $L_2$ that contribute to the four-point correlation function $\langle \phi_+^2 (t)\chi_+(t_1)\chi_+(t_2)\rangle_c $ at order $O(g^2)$. The $\bigoplus$ denotes the composite operator $\phi^2_+(t)$.}
	\label{fig:loop-diag}
\end{figure}

We commence by first computing $\langle \phi_+^2 (t)\chi_+(t_1)\chi_+(t_2)\rangle_c $. At order $\mathcal{O}(g^2)$, it receives contributions from the one-loop diagrams $L_1$ and $L_2$ shown in Fig.~\ref{fig:loop-diag}. Two additional diagrams, $L_1^\prime$ and $L_2^\prime$ which differ from $L_1$ and $L_2$ only with respect to the exchange  $ \vec{p}_1 \leftrightarrow \vec{p}_2$ also contribute to this correlation function. A detailed description of the loop calculations has been summarized in appendix \ref{sec:loop-calc}. In the calculations shown below, we delve into the explicit details for only $\langle \phi_+^2 (t)\chi_+(t_1)\chi_+(t_2)\rangle_c $, for the other 15 correlation functions $\langle \phi_i (t)\phi_j (t)\chi_k(t_1)\chi_l(t_2)\rangle_c $ similar steps must be followed. It must be emphasized that each of  the other 15 correlation functions can be calculated using diagrams similar to $L_1$, $L_2$ (along with their momentum-exchanged ($ \vec{p}_1 \leftrightarrow \vec{p}_2$) counterparts).
\\
\subsubsection*{\centering\underline{Case 1:$\;\; t_0<t_{in}\leq t < t_1 <t_2$}} 

\begin{figure}[h]
	\centering
	\includegraphics[width=0.7\linewidth]{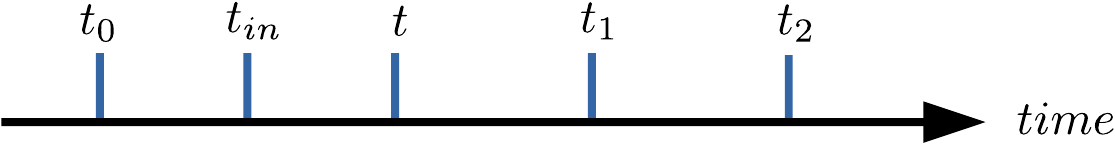}
	\caption{\sf The chronology of events in the first case. Here, $t_0$ corresponds to the time when mass quench occurs, $t_{in}$ indicates the time when the interaction between $\phi$ and $\chi$ fields is turned on and $t$ refers to the time of the measurement. $t_1$, $t_2$ correspond to the external states.}
	\label{fig:time1}
\end{figure}

The leading order term on the right hand side for $\langle \phi_+^2 (t)\chi_+(t_1)\chi_+(t_2)\rangle_c $ in this case, obtained after evaluating the time integrals in the calculation of loop diagrams, see appendix \ref{sec:loop-calc}, is comprised of an integrand of order $1/|\vec{k}|^3$. We can identify four-point correlation functions of $\chi$ in the result, each multiplied by a time-dependent coefficient as shown below: 
{\small\begin{eqnarray}\label{eq:case1-4pt-chi-corr}
		L_1 + L_2 + L_1^\prime + L_2^\prime &= &\Bigg( \Big[ \frac{e^{i \Omega_{p_1}(t_1-t_{in})}}{2\Omega_{p_1}}\frac{e^{i \Omega_{p_2}(t_2-t_{in})}}{2\Omega_{p_2}} \Big]\, c_1 (t_{in},\,t_0,\,t) \,+\,
		\nonumber\\ 
		& & 
		\Big[ \frac{e^{-i \Omega_{p_1}(t_1-t_{in})}}{2\Omega_{p_1}}\frac{e^{-i \Omega_{p_2}(t_2-t_{in})}}{2\Omega_{p_2}} \Big]\, c_2 (t_{in},\,t_0,\,t) \,+\, \nonumber\\ 
		& & \Big[ \frac{e^{i \Omega_{p_1}(t-t_1)}}{2\Omega_{p_1}}\frac{e^{i \Omega_{p_2}(t-t_2)}}{2\Omega_{p_2}} \Big]\, c_3 (t_{in},\,t_0,\,t) \Bigg)+ \vec{p}_1 \rightarrow \vec{p}_2\;,
\end{eqnarray}}
which implies the following:
{\small\begin{eqnarray}
		\langle \phi_+^2 (t)\chi_+(t_1)\chi_+(t_2)\rangle_c &=&  \langle \chi_-^2(t_{in}) \chi_+(t_1) \chi_+(t_2) \rangle_c \,  c_1 (t_{in},\,t_0,\,t) 
		\nonumber\\
		&\,+\, &
		\langle \chi_+^2(t_{in}) \chi_+(t_1) \chi_+(t_2) \rangle_c \,  c_2 (t_{in},\,t_0,\,t)  \nonumber\\
		& \,+\,&  \langle \chi_+^2(t) \chi_+(t_1) \chi_+(t_2) \rangle_c \,  c_3 (t_{in},\,t_0,\,t). 
\end{eqnarray}}
Here, the connected four-point correlation functions of $\chi$ have been identified as:
{\small\begin{eqnarray}\label{eq:chi-4-pt-correlator}
		\langle \chi_-^2(t_{in}) \chi_+(t_1) \chi_+(t_2) \rangle_c &=& \frac{e^{i \Omega_{p_1}(t_1-t_{in})}}{2\Omega_{p_1}}\frac{e^{i \Omega_{p_2}(t_2-t_{in})}}{2\Omega_{p_2}} + (\vec{p}_1 \rightarrow \vec{p}_2),\nonumber\\
		\langle \chi_+^2(t_{in}) \chi_+(t_1) \chi_+(t_2) \rangle_c &=& \frac{e^{-i \Omega_{p_1}(t_1-t_{in})}}{2\Omega_{p_1}}\frac{e^{-i \Omega_{p_2}(t_2-t_{in})}}{2\Omega_{p_2}} + (\vec{p}_1 \rightarrow \vec{p}_2),\nonumber\\
		\langle \chi_+^2(t) \chi_+(t_1) \chi_+(t_2) \rangle_c  &=& \frac{e^{i \Omega_{p_1}(t-t_1)}}{2\Omega_{p_1}}\frac{e^{i \Omega_{p_2}(t-t_2)}}{2\Omega_{p_2}}  + (\vec{p}_1 \rightarrow \vec{p}_2).
\end{eqnarray}}
The time-dependent coefficients are each, in fact, integrals over the momentum $\vec{k}$, with the integrands being functions of $\omega_k$ and $\omega_{0k}$. Absorbing the coupling constant $g^2$ within these coefficients allows us to write them as:
\begin{eqnarray}\label{cs}
		c_1 (t_{in},\,t_0,\,t) &=& \frac{g^2}{8} \int\frac{d^3\vec{k}}{(2\pi)^3} \frac{1}{\omega_k^2 \omega_{0k}}\,\big[\cos\omega_k(t-t_0) + i \frac{\omega_{0k}}{\omega_k} \sin\omega_k(t-t_0)\big]^2
		\times \nonumber\\
		& &  \qquad \qquad \big[\cos 2\omega_k(t_{in}-t_0) - i \frac{(\omega_k^2 + \omega_{0k}^2)}{2\omega_k\omega_{0k}}\sin 2\omega_k(t_{in}-t_0) \big], \nonumber \\
		c_2 (t_{in},\,t_0,\,t) &=& \frac{g^2}{8} \int\frac{d^3\vec{k}}{(2\pi)^3} \frac{1}{\omega_k^2 \omega_{0k}}
		\big[\cos\omega_k(t-t_0) - i \frac{\omega_{0k}}{\omega_k} \sin\omega_k(t-t_0)\big]^2 \times \nonumber\\
		& &  \qquad \qquad \big[\cos 2\omega_k(t_{in}-t_0) + i \frac{(\omega_k^2 + \omega_{0k}^2)}{2\omega_k\omega_{0k}}\sin 2\omega_k(t_{in}-t_0) \big],  \nonumber \\
		c_3 (t_0,\,t) &=& -\frac{g^2}{8} \int\frac{d^3\vec{k}}{(2\pi)^3} \frac{1}{\omega_k^2 \omega_{0k}} \Big[\big(\cos\omega_k(t-t_0)+i\frac{\omega_{0k}}{\omega_k}\sin\omega_k(t-t_0)\big)^2  \times \nonumber\\
		& & \qquad \qquad \big(\cos 2\omega_k(t-t_0) - i \frac{(\omega_k^2 + \omega_{0k}^2)}{2\omega_k\omega_{0k}}\sin 2\omega_k(t-t_0) \big)
		+ \nonumber\\
		& & \qquad \qquad \big(\cos\omega_k(t-t_0)-i\frac{\omega_{0k}}{\omega_k}\sin\omega_k(t-t_0)\big)^2 \times \nonumber\\
		& & \qquad \qquad \big(\cos 2\omega_k(t-t_0) + i \frac{(\omega_k^2 + \omega_{0k}^2)}{2\omega_k\omega_{0k}}\sin 2\omega_k(t-t_0) \big)\Big].
\end{eqnarray}

In the limit of no mass quench, i.e., for $m_0 \rightarrow m$, we , these simplify considerably:
{\small\begin{eqnarray}
	c_1 (t_{in},\,t_0,\,t) 
	& \rightarrow & \frac{g^2}{8} \int\frac{d^3\vec{k}}{(2\pi)^3} \frac{1}{\omega_k^3}e^{2 i \omega_k(t-t_{in})},  \hspace{1cm}
	c_3 (t_0,\,t) 
	 \rightarrow  -\frac{2 g^2}{8} \int\frac{d^3\vec{k}}{(2\pi)^3} \frac{1}{\omega_k^3}, \nonumber\\
	c_2 (t_{in},\,t_0,\,t) 
	&\rightarrow&  \frac{g^2}{8} \int\frac{d^3\vec{k}}{(2\pi)^3} \frac{1}{\omega_k^3}e^{-2 i \omega_k(t-t_{in})}.
	\end{eqnarray}}
After repeating similar steps for all 16 correlation functions $\langle \phi_i (t)\phi_j(t)\chi_k(t_1)\chi_l(t_2)\rangle_c$ with $i, j, k, l \in \{+, - \}$, summing them and setting $\phi_+ = \phi_- = \phi$, as well as $\chi_+ = \chi_- = \chi$ to return to a description in terms of the physical fields $\phi$ and $\chi$ \cite{Dresti:2013kya}, we find that%
{\small\begin{eqnarray}
\langle \phi^2(t) \chi(t_1)\chi(t_2)\rangle_c &=&\underbrace{\big( c_1 (t_{in},\,t_0,\,t)+ c_2(t_{in},\,t_0,\,t) \big)}_{K_1(t-t_{in})} \langle \chi^2(t_{in}) \chi(t_1)\chi(t_2) \rangle_c \nonumber \\ &&\qquad+\,\,  \underbrace{c_3 (t_0,\,t)}_{c_3(t-t_0)} \, \langle \chi^2(t) \chi(t_1)\chi_l(t_2) \rangle_c .
\end{eqnarray}}
The non-local and local kernels have been identified as, ignoring terms of $ \mathcal{O}(1/|\vec{k}|^5) $:
{\small\begin{eqnarray}\label{K1}
K_1(t-t_{in}) &=& c_1 (t_{in},\,t_0,\,t)+ c_2(t_{in},\,t_0,\,t)\nonumber\\
&\approx& \frac{g^2}{8}\int \frac{d^3\vec{k}}{(2\pi)^3}\frac{\omega_k^2+\omega_{0k}^2}{\omega_k^4\omega_{0k}} \cos(2\omega_k(t-t_{in})),
\end{eqnarray}}
\vspace{-0.5cm}{\small\begin{eqnarray}\label{c3} 
\hspace{-2cm}c_3 &\approx& -\frac{g^2}{8} \int \frac{d^3\vec{k}}{(2\pi)^3}\frac{\omega_k^2+\omega_{0k}^2}{\omega_k^4\,\omega_{0k}}. 
\end{eqnarray}}
The relation between the four-point correlators allows us to write  the following relation between composite operators in terms of (non-)local kernels:
{\small\begin{eqnarray}
	\phi^2_R(t)= \phi^2(t) +K_{1}(t-t_{in}) \chi^2(t_{in}) + c_3 \chi^2(t) + \text{counter terms}. \label{eq:case1-phi2}
	\end{eqnarray}}
In the above, $\phi^2(t)$ arises from the disconnected diagram, where $\phi(t)$ is the bare field. The counter terms are necessary to renormalize the divergence within $c_3$. Details pertaining to the renormalization of $c_3$ as well the features of $K_1$ have been discussed in \S \ref{sec:analysis}.

\subsubsection*{\centering\underline{Case 2:$\,\,t_{in}<t_0< t < t_1 <t_2$}} 

\begin{figure}[h]
	\centering
	\includegraphics[width=0.7\linewidth]{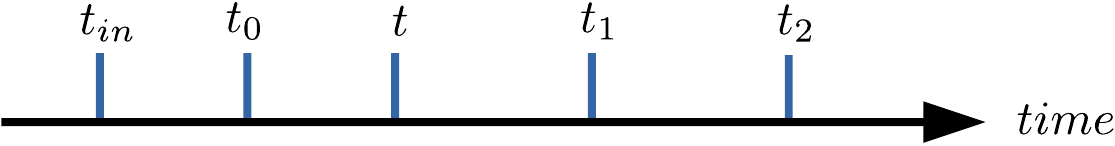}
	\caption{\sf The chronology of events in the second case. Once again, $t_0$ corresponds to the time when mass quench occurs, $t_{in}$  indicates the time when the interaction between $\phi$ and $\chi$ fields is turned on and $t$ refers to the time of the measurement. $t_1$, $t_2$ correspond to the external states.}
	\label{fig:time2}
\end{figure}

The leading order term on the right hand side for $\langle \phi_+^2 (t)\chi_+(t_1)\chi_+(t_2)\rangle_c $, obtained after evaluating the time integrals in the calculation of loop diagrams, is once again comprised of integrands of order $1/|\vec{k}|^3$. We proceed in a manner similar to Case 1 by expressing the result as a combination of four-point correlation functions of $\chi$ and by studying the behaviour of their time-dependent coefficients. 

{\small\begin{eqnarray}\label{eq:case2-4pt-chi-corr}
	L_1 + L_2 + L_1^\prime + L_2^\prime &= &\Bigg( \Big[ \frac{e^{i \Omega_{p_1}(t_1-t_{in})}}{2\Omega_{p_1}}\frac{e^{i \Omega_{p_2}(t_2-t_{in})}}{2\Omega_{p_2}} \Big]\,\, c_1^\prime (t_{in},\,t_0,\,t) \,+\,
	\nonumber\\ 
	& & 
	\Big[ \frac{e^{-i \Omega_{p_1}(t_1-t_{in})}}{2\Omega_{p_1}}\frac{e^{-i \Omega_{p_2}(t_2-t_{in})}}{2\Omega_{p_2}} \Big]\, \,c_2^\prime (t_{in},\,t_0,\,t) \,+\, \nonumber\\ 
	& & \Big[ \frac{e^{i \Omega_{p_1}(t-t_1)}}{2\Omega_{p_1}}\frac{e^{i \Omega_{p_2}(t-t_2)}}{2\Omega_{p_2}} \Big]\,\, c_3 ^\prime(t_{in},\,t_0,\,t) \Bigg)+ \vec{p}_1 \rightarrow \vec{p}_2\;,
	\end{eqnarray}}
which implies the following:
{\small\begin{eqnarray}
	\langle \phi_+^2 (t)\chi_+(t_1)\chi_+(t_2)\rangle_c &=&  \langle \chi_-^2(t_{in}) \chi_+(t_1) \chi_+(t_2) \rangle_c \,\,  c_1^\prime (t_{in},\,t_0,\,t) 
	\nonumber\\
	&\,+\, &
	\langle \chi_+^2(t_{in}) \chi_+(t_1) \chi_+(t_2) \rangle_c \,\,  c_2^\prime (t_{in},\,t_0,\,t)  \nonumber\\
	& \,+\,&  \langle \chi_+^2(t) \chi_+(t_1) \chi_+(t_2) \rangle_c \,\,  c_3 ^\prime(t_{in},\,t_0,\,t). 
	\end{eqnarray}}
The four-point correlation functions of $\chi$ above are the same as the ones defined in Eq.~\eqref{eq:chi-4-pt-correlator}. The difference between Case 1 and 2 appears within the time-dependent coefficients which for Case 2 are written below:

{\small\begin{eqnarray}\label{cps}
		c^\prime_1 (t_{in},t_0,t) &=& \frac{g^2}{8} \int\frac{d^3\vec{k}}{(2\pi)^3} \frac{1}{\omega_{0k}^3}
		e^{2i \omega_{0k}(t_0-t_{in})}
		\big[\cos\omega_k(t-t_0) + i \frac{\omega_{0k}}{\omega_k} \sin\omega_k(t-t_0)\big]^2   \nonumber\\
		c^\prime_2 (t_{in},t_0,t) &=& \frac{g^2}{8} \int\frac{d^3\vec{k}}{(2\pi)^3} \frac{1}{\omega_{0k}^3}
		e^{-2i \omega_{0k}(t_0-t_{in})}
		\big[\cos\omega_k(t-t_0) - i \frac{\omega_{0k}}{\omega_k} \sin\omega_k(t-t_0)\big]^2 \nonumber \\
		c^\prime_3 (t_0,t) &=& -\frac{g^2}{8} \int\frac{d^3\vec{k}}{(2\pi)^3} \frac{1}{\omega_k^2 \omega_{0k}} \Big[\big(\cos\omega_k(t-t_0)+i\frac{\omega_{0k}}{\omega_k}\sin\omega_k(t-t_0)\big)^2  \times \nonumber\\
		& & \qquad \qquad \big(\cos 2\omega_k(t-t_0) - i \frac{(\omega_k^2 + \omega_{0k}^2)}{2\,\omega_k\,\omega_{0k}}\sin 2\omega_k(t-t_0) \big)
		+ \nonumber\\
		& & \qquad \qquad \big(\cos\omega_k(t-t_0)-i\frac{\omega_{0k}}{\omega_k}\sin\omega_k(t-t_0)\big)^2 \times \nonumber\\
		& &\qquad \qquad \big(\cos 2\omega_k(t-t_0) + i \frac{(\omega_k^2 + \omega_{0k}^2)}{2\,\omega_k\,\omega_{0k}}\sin 2\omega_k(t-t_0) \big)\Big]\;. 
\end{eqnarray}}
In the limit of no mass quench, i.e., for $m_0 \rightarrow m$, we obtain
{\small\begin{eqnarray}
		c^\prime_1 (t_{in},t_0,t) & \rightarrow & \frac{g^2}{8} \int\frac{d^3\vec{k}}{(2\pi)^3} \frac{1}{\omega_k^3}e^{2 i \omega_k(t-t_{in})}\;,  \qquad\qquad c^\prime_3 (t_0,t)  \rightarrow  -\frac{2g^2}{8} \int\frac{d^3\vec{k}}{(2\pi)^3} \frac{1}{\omega_k^3}\;, \nonumber\\
		c^\prime_2 (t_{in},t_0,t) & \rightarrow & \frac{g^2}{8} \int\frac{d^3\vec{k}}{(2\pi)^3} \frac{1}{\omega_k^3}e^{-2 i \omega_k(t-t_{in})}\;. 	
\end{eqnarray}}
After evaluating all 16 correlation functions $\langle \phi_i (t)\phi_j(t)\chi_k(t_1)\chi_l(t_2)\rangle_c$ with $i, j, k, l \in \{+, - \}$, summing them and setting $\phi_+ = \phi_- = \phi$, as well as $\chi_+ = \chi_- = \chi$ to return to a description in terms of the physical fields $\phi$ and $\chi$ \cite{Dresti:2013kya}, we can identify the time-dependent kernel: 

{\small\begin{eqnarray}
\langle \phi^2(t) \chi(t_1)\chi(t_2)\rangle_c &=& \underbrace{\big( c^\prime_1 (t_{in},\,t_0,\,t)+ c^\prime_2(t_{in},\,t_0,\,t) \big)}_{K_2(t-t_0;\,t_0-t_{in})} \langle \chi^2(t_{in}) \chi(t_1)\chi(t_2) \rangle_c \nonumber \\
&&+  \underbrace{c^\prime_3 (t_0,\,t)}_{c_3^\prime(t-t_0)} \, \langle \chi^2(t) \chi(t_1)\chi_l(t_2) \rangle_c \;,
\end{eqnarray}} 
(ignoring terms of $ \mathcal{O}(1/|\vec{k}|^5) $), we can identify the non-local and local kernels as:	
{\small\begin{eqnarray}\label{K2} 
K_2(t-t_0;\,t_0-t_{in}) &=& c^\prime_1 (t_{in},\,t_0,\,t)+ c^\prime_2(t_{in},\,t_0,\,t) \nonumber \\
&\approx& \frac{g^2}{8} \int\frac{d^3\vec{k}}{(2\pi)^3} \Bigg[ \frac{\omega_k^2+\omega_{0k}^2}{\omega_k^2\omega_{0k}^3} \cos(2\omega_{0k}(t_0-t_{in})) \cos(2\omega_k(t-t_0)) \nonumber\\ 
& & \qquad - \frac{1}{\omega_k\omega_{0k}^2} 2\sin(2\omega_{0k}(t_0-t_{in})) \sin(2\omega_k(t-t_0)) \Bigg],
\end{eqnarray}}
\vspace{-0.5cm}{\small\begin{eqnarray}\label{c3p}
\hspace{-3.7cm}c_3^\prime&\approx& -\frac{g^2}{8} \int\frac{d^3\vec{k}}{(2\pi)^3} \frac{\omega_k^2+\omega_{0k}^2}{\omega_k^4\omega_{0k}}\;.
\end{eqnarray}}
Once again, the relation between the four-point correlators informs the mixing of operators weighted by local and non-local kernels:
{\small\begin{eqnarray}
	\phi^2_R(t)= \phi^2(t) + K_2(t-t_0;\,t_0-t_{in}) \chi^2(t_{in}) + c_3^\prime \chi^2(t) + \text{counter terms}.\label{eq:case2-phi2}
	\end{eqnarray}}
As before, $\phi^2(t)$ arises from the disconnected diagram, with $\phi(t)$ being the bare field and counter terms are necessary to renormalize the divergence within $c_3^\prime$. The characteristics of $K_2$ and details pertaining to the renormalization of $c_3^\prime$ have been discussed in \S \ref{sec:analysis}.

\subsubsection*{\centering\underline{Comparison of the two cases}} 
\vspace{0.3cm}
Using the expressions for $K_1(t-t_{in})$, $c_3$, $K_2(t-t_0;\,t_0-t_{in})$ and $c_3^\prime$, we can conduct a straightforward comparison of the two cases while imposing various limits on $t_0$, $t_{in}$ and $t$. However, before doing so, the following points must be emphasized:

\begin{enumerate}
	\item In case 1, the ordering $t_0 < t_{in} <  t$ holds. Therefore, while we can take the limits $t \rightarrow t_{in}$ and $t_0 \rightarrow t_{in}$ separately, setting $t\rightarrow t_0$ will automatically imply that all three events, i.e., mass quench, interaction quench and the measurement are occurring at the same instant. 
	  
	 \item On the other hand, for case 2, we have the ordering $t_{in} < t_0 <  t$. As a consequence, we can not take the limit $t \rightarrow t_{in}$ directly, i.e., the limits $t \rightarrow t_0$ and $t_{in} \rightarrow t_0$ must be taken separately to describe the simultaneous occurrence of the three events.
\end{enumerate}
Keeping the above points in mind, we can deduce relations between the local and non-local kernels and also obtain simplified forms for them.

\begin{enumerate}
	\item For case 1, it is clear from Eqs.~\eqref{K1} and \eqref{c3}, that there exists a direct relation between $c_3$ and $K_1$ when we set $t\rightarrow t_{in}$ in case 1, i.e., $c_3 = - K_1(0)$. As a consequence of this, no operator mixing occurs 
	
	\vspace{-0.6cm}
	{\small\begin{eqnarray}
			\langle \phi^2(t) \chi(t_1)\chi(t_2)\rangle_c &=& \Big( K_1(t_{in}-t_{in}) +  c_3 \Big)\, \langle \chi^2(t_{in}) \chi(t_1)\chi(t_2)\rangle_c  \nonumber\\
			&=& \Big( K_1(0) -  K_1 (0) \Big)\, \langle \chi^2(t_{in}) \chi(t_1)\chi(t_2)\rangle_c \nonumber\\
			&=& 0.
	\end{eqnarray}} 
	This property holds irrespective of whether mass quench has occurred or not, even though $K_1(0)$ still contains a signature of the mass quench on account of the integrand being a function of both $\omega_k$ and $\omega_{0k}$.

	\item For case 2, while we cannot directly set $t \rightarrow t_{in}$, even if we inspect the situation where $t \rightarrow t_0$ and $t_{in} \rightarrow t_0$ hold simultaneously, we notice that $K_2(0,0) + c_3^\prime \neq 0$. This turns into equality only when $\omega_k =\omega_{0k}$ is also enforced, i.e., when no mass quench occurs.
	
	\item In the event of no mass quench, i.e, $m_0 = m$ and consequently $\omega_k =\omega_{0k}$, the results of \cite{Dresti:2013kya} are reproduced for both cases:
	{\small\begin{eqnarray}
			K_1(t-t_{in})&& \xrightarrow{\omega_k\, =\, \omega_{0k}} K(t-t_{in}), \quad \quad c_3 \xrightarrow{\omega_k\, =\, \omega_{0k}} -K(0), \nonumber\\
			K_2(t-t_0;t_0-t_{in}) && \xrightarrow{\omega_k\, =\, \omega_{0k}} K(t-t_{in}), \quad\quad c_3^\prime \xrightarrow{\omega_k\, =\, \omega_{0k}} -K(0).
	\end{eqnarray}}
\noindent	where time-dependent kernel is
	{\small\begin{eqnarray}
			K(t-t_{in}) = \cfrac{g^2}{4} \int\cfrac{d^3\vec{k}}{(2\pi)^3} \cfrac{\cos(2 \omega_k (t- t_{in}))}{\omega_k^3}\;.
	\end{eqnarray}}
\end{enumerate}

\section{Analysis of (non-)local kernels}\label{sec:analysis}

In this section we evaluate the local as well as non-local kernels, discussed in \S \ref{sec:operator-mixing}, that facilitate operator mixing. 
\subsection{Numerical estimation of the kernels}

\begin{figure}[h]
	\centering
	\includegraphics[scale=0.6]{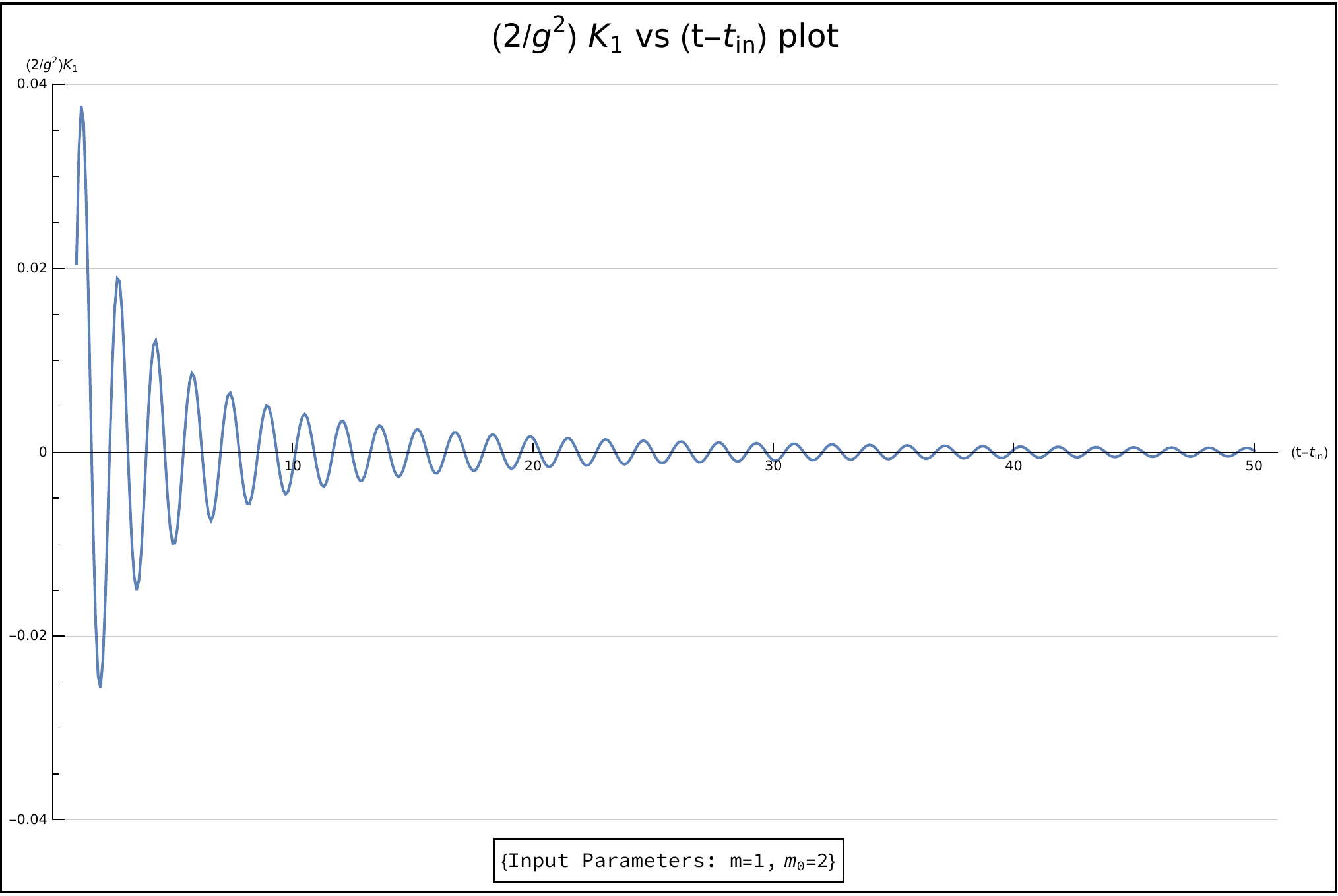}
	\caption{\sf Plot displaying $(2/g^2)\,K_1$ as a function of $(t-t_{in})$. The parameters $m_0$ and $m$ have been fixed at constant values. It can be seen that $K_1 $ vanishes at later time.}
	\label{fig:k1plot}
\end{figure}

\begin{figure}[h]
	\centering
	\includegraphics[scale=0.6]{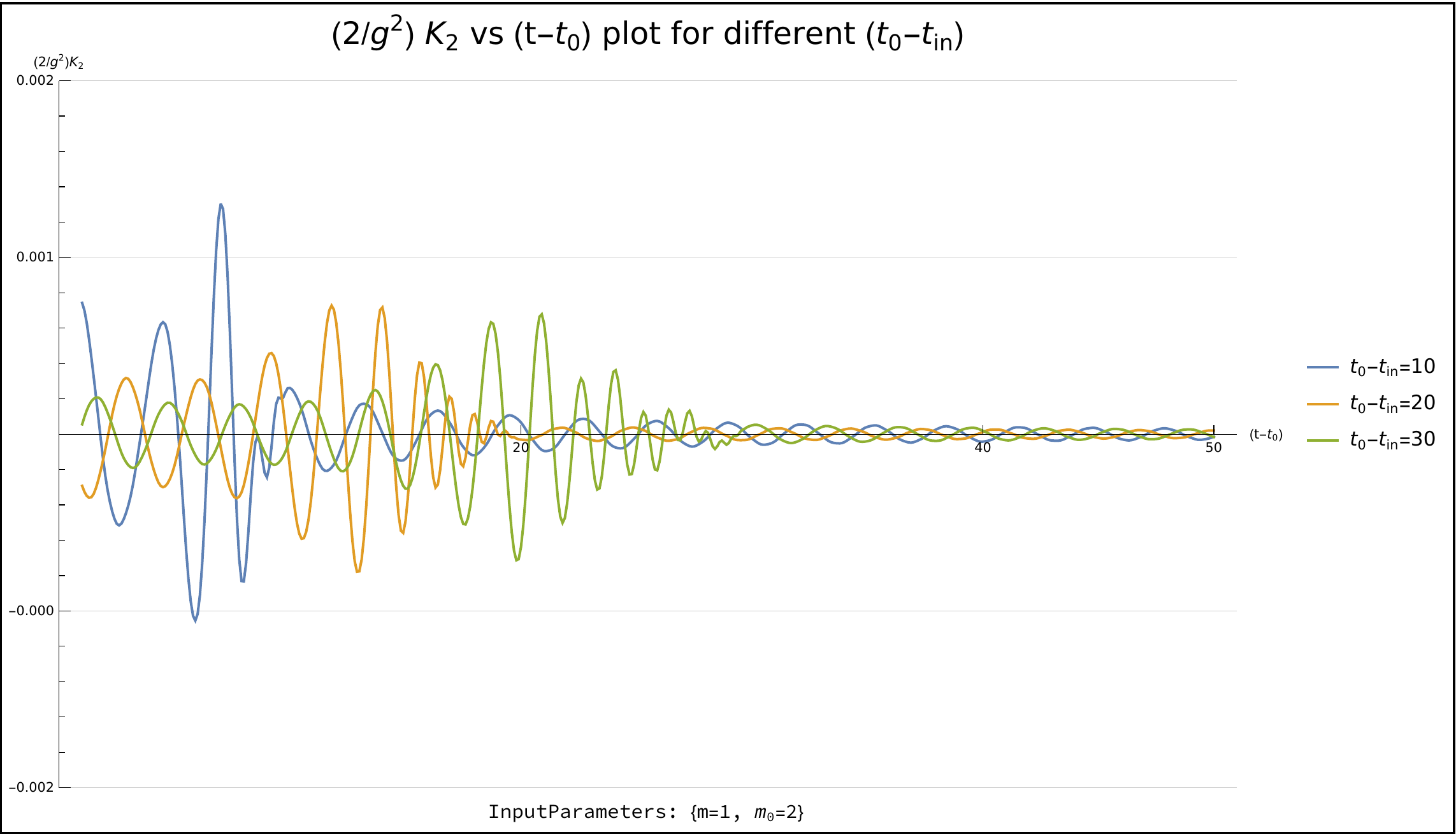}
	\caption{\sf Plot displaying $(2/g^2)\,K_2$ as a function of $(t-t_0)$ for three different choices of the interval $(t_0-t_{in})$. The parameters $m_0$ and $m$ have been fixed at constant values. Similar to $K_1$, the kernel $K_2$ also vanishes at a later time.}
	\label{fig:k2plot1}
\end{figure}

Both $K_1$ and $K_2$, see Eqs.~\eqref{K1} and \eqref{K2}, contain oscillatory functions in the integrand, weighted by momentum dependent factors. Computing these integrals in a closed form is difficult. In \cite{Dresti:2013kya}, the kernel had been identified in terms of Meijer-G functions. In appendix \ref{sec:kernel-series-expansion}, we have done a similar identification in terms of Meijer-G functions after doing a Taylor series expansion of the integrand assuming a small mass quench. In this section, on the other hand, we highlight the general features of $K_1$ and $K_2$ by evaluating the integrals numerically. A graphical demonstration of $K_1$ (scaled down by $g^2/2$) as a function of $t-t_{in}$ has been provided in Fig.~\ref{fig:k1plot} and highlights a decaying and oscillating profile. Similarly, 
$K_2$ (scaled down by $g^2/2$) as a function of $t-t_0$ for fixed choices of the interval $t_0-t_{in}$ has been displayed in Fig.~\ref{fig:k2plot1}. Once again, one can notice oscillatory behaviour. The different choices of $t_0-t_{in}$ influence the location of the peak, but ultimately each of the profiles decays for large-($t-t_0$). 

\subsection{Renormalization of divergent integrals}

\begin{figure}[h]
	\centering
	\includegraphics[scale=0.7]{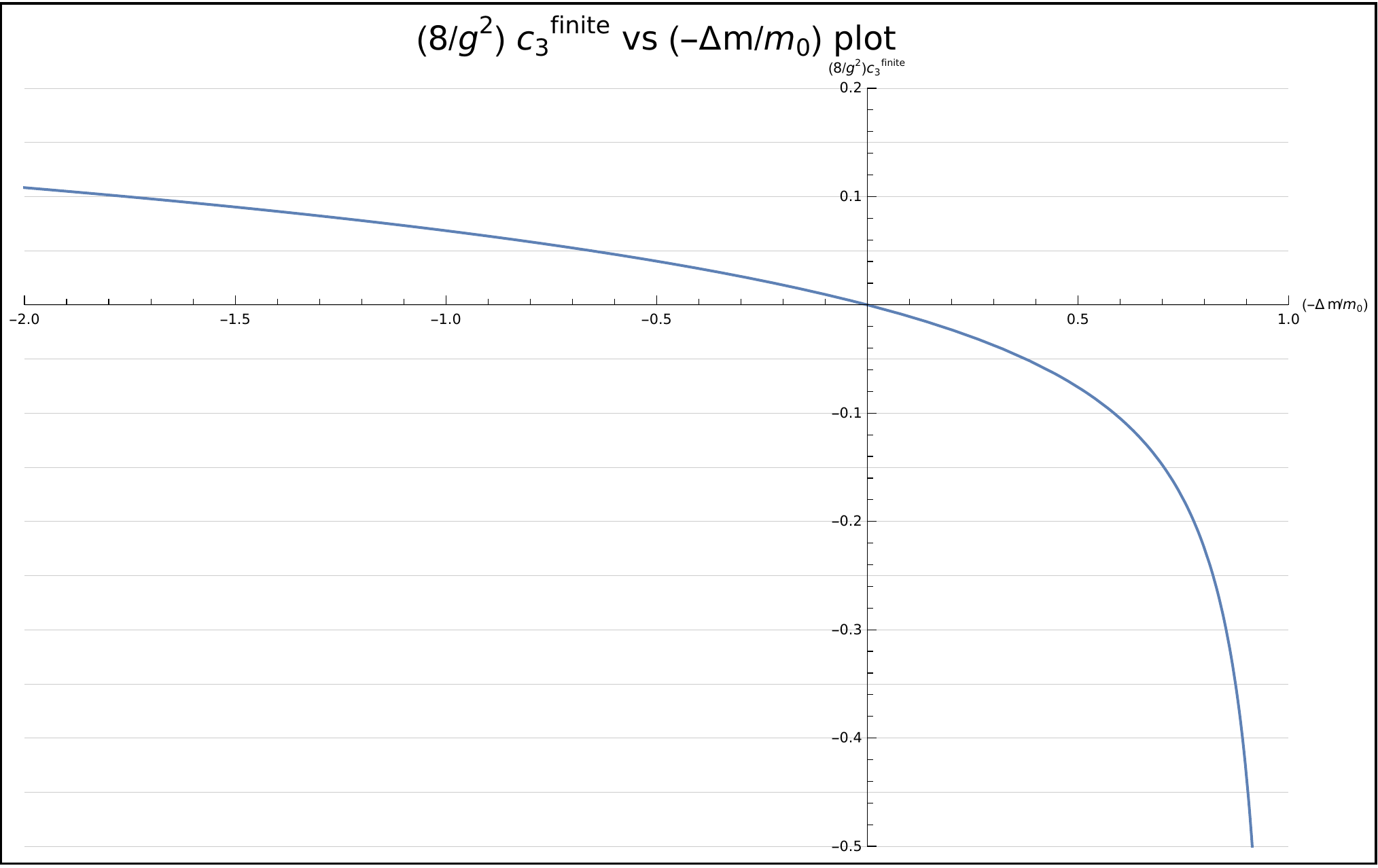}
	\caption{\sf Plot displaying $(8/g^2)\,c_3^\text{finite}$ as function of $x = -\Delta m/m_0$.}
	\label{fig:c3plot}
\end{figure}

From Eqs.~\eqref{c3} and \eqref{c3p}, it is evident that $c_3 = c_3'$. So, their features can be studied together. The integration over $\vec{k}$ generates a logarithmic ultraviolet divergence that along with the appropriate counter-terms, see Eqs.~\eqref{eq:case1-phi2}, \eqref{eq:case2-phi2}, renormalizes the composite operator in both the cases. Here, $c_3$ and $c_3^\prime$ contain divergent and finite parts. The result of evaluating the $k$-integral in the the expression for $c_3$ (and $c_3^\prime$), which we will denote as $\widetilde{c}_3$ can be split based on whether $m > m_0$ or $m < m_0$ as follows: 
{\small\begin{eqnarray}
		\widetilde{c}_3 &=& -\cfrac{g^2}{16\pi^2}\Bigg[\log \left(\cfrac{\Lambda^2}{m_0^2}\right)+\log4-\cfrac{1}{2} + \cfrac{1}{2 \sqrt{m_0^2 - m^2}} \, \left(\cfrac{m_0^2 - 4 m^2}{m}\right)\cos^{-1}\left(\cfrac{m}{m_0}\right)\,\Bigg], \quad\text{for } m < m_0; \nonumber\\
		&=& -\cfrac{g^2}{16\pi^2}\Bigg[\log \left(\cfrac{\Lambda^2}{m_0^2}\right)+\log4-\cfrac{1}{2} + \cfrac{1}{2 \sqrt{m^2 - m_0^2}} \, \left(\cfrac{m_0^2 - 4 m^2}{m}\right)\cosh^{-1}\left(\cfrac{m}{m_0}\right)\,\Bigg], \quad\text{for } m > m_0.\nonumber\\
\end{eqnarray}}
Here $\Lambda$ denotes the large momentum cut-off. From this complete expression, we choose the counter-terms (mimicking $\overline{MS}$ scheme) judiciously by absorbing the divergent as well as  the universal constant terms to ensure that operator mixing does not occur in the no mass quench limit, i.e., 
{\small\begin{eqnarray}
	c_3^{\text{CT}} \,=\, \widetilde{c}_3\,(m\rightarrow m_0) \,=\,  -\cfrac{g^2}{16\pi^2}\Bigg[\log \left(\cfrac{\Lambda^2}{m_0^2}\right)+ \log4-2\,\Bigg].
	\end{eqnarray}}
Next, the finite part of $\widetilde{c}_3$ can be identified as $c_3^{\text{finite}}$ = $\widetilde{c}_3 - c_3^{\text{CT}}$ and the explicit form can be written as:
{\small\begin{eqnarray}\label{eq:c3-finite}
	c_3^{\text{finite}} &=& -\cfrac{g^2}{16 \pi^2}\,\left[ \cfrac{1}{2 \sqrt{m_0^2 - m^2}} \, \left(\cfrac{m_0^2 - 4 m^2}{m}\right)\cos^{-1}\left(\cfrac{m}{m_0}\right)+\cfrac{3}{2}\;\right], \quad\text{for } m < m_0; \nonumber\\
	&=& -\cfrac{g^2}{16 \pi^2}\,\left[ \cfrac{1}{2 \sqrt{m^2 - m_0^2}} \, \left(\cfrac{m_0^2 - 4 m^2}{m}\right)\cosh^{-1}\left(\cfrac{m}{m_0}\right)+\cfrac{3}{2}\;\right], \hspace{0.1cm}\text{for } m > m_0.
	\end{eqnarray}}
If we denote the change in mass after quench by $\Delta m$, i.e., if $ m=m_0 + \Delta m $, then by defining a dimensionless parameter $ x= -\Delta m/m_0 $ and substituting $ m=m_0\,(1-x) $ in Eq.~\eqref{eq:c3-finite} we obtain,%
{\small\begin{eqnarray}\label{eq:c3-finite-x}
	c_3^{\text{finite}} &=& -\cfrac{g^2}{16\pi^2}\,\left[ \cfrac{1}{2\sqrt{x(2-x)}} \, \left(\cfrac{3-8x+4x^2}{(x-1)}\right)\cos^{-1}\left(1-x\right)+\cfrac{3}{2}\;\right], \quad\text{for } m < m_0\;; \nonumber \\
	&=& -\cfrac{g^2}{16\pi^2}\,\left[\cfrac{1}{2 \sqrt{x(x-2)}} \, \left(\cfrac{3-8x+4x^2}{(x-1)}\right)\cosh^{-1}\left(1-x\right)+\cfrac{3}{2}\;\right], \;\text{for } m > m_0\;.
	\end{eqnarray}}
\noindent
It can be seen clearly that the condition $ (x < 1) $ must hold, otherwise $ m $ will become negative. A plot of $c^{\text{finite}}_3$ as a function of $x$ has been shown in Fig.~\ref{fig:c3plot}. It can be seen from the figure that $c_3^\text{finite}$ becomes largely negative as $ x \rightarrow 1 $, i.e., in the limit of vanishing $ m $. The curve passing through the origin signifies the absence of operator mixing at no mass quench limit. 

\section{Implications of operator mixing for the $\chi$ potential}\label{sec:chi-potential}

In this section, we demonstrate how the operator mixing of one field can affect the potential of  other one for a two scalar field simplified system. Though the impact is mutual, we highlight the deformation of  $\chi$ potential due to the $\phi^2(t)$ operator mixing. It must be mentioned that initially the  $\chi$ potential is parabolic since $M^2, \lambda_\chi >0$. We investigate, in this section, whether the $\phi^2(t)$ operator mixing can flip the sign of the $\chi^2$ term, leading to a wine-bottle shaped potential for $\chi$. This would imply a  hint of  possible phase transition or spontaneous breaking of any underlying symmetry, induced by the operator mixing, as an artifact of mass and interaction quenches. It must be emphasized that phase transitions of this nature are intricate as one is no longer limited to ground state physics. 

\begin{figure}
	\centering
	\includegraphics[scale=0.45]{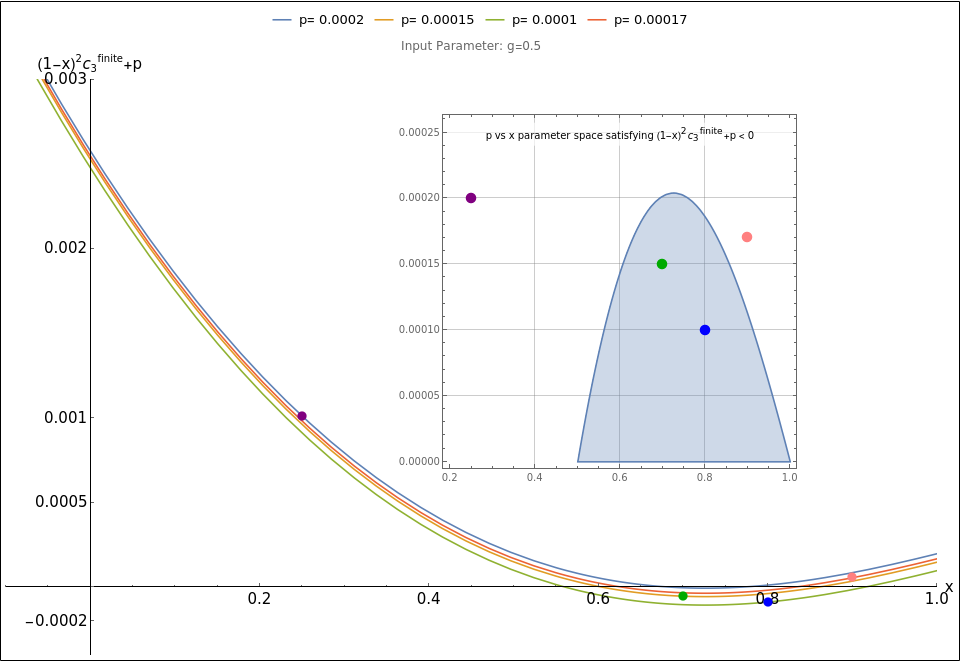}
	\caption{\sf Plot of $[ (1-x)^2 c_3^{\text{finite}} + p ]$ as a function of $x$ and for some fixed values of $ p $. The inset figure highlights the ($x,p$) parameter space that allows for negative $ a_1 $. The points inside the shaded region correspond to potentials with the shape of a \textit{wine-bottle}, those lying outside correspond to parabolic potential. Points chosen from the inset have been suitably highlighted on the main plot.}
	\label{fig:solnplt}
\end{figure}

\begin{figure}[htbp!]
	\centering
	\includegraphics[width=0.76\linewidth]{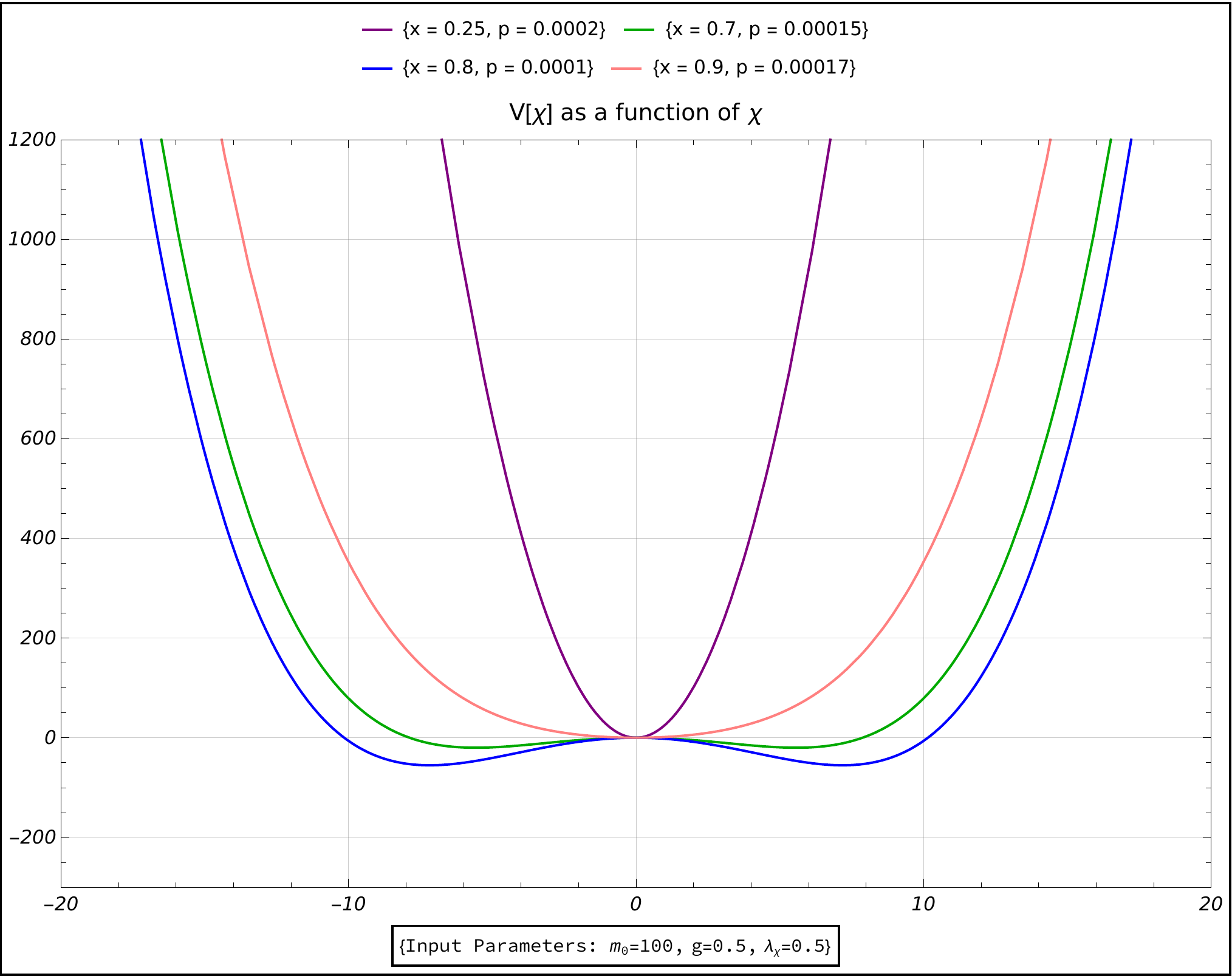}
	\caption{\sf Plot of $V(\chi)$ as a function of $\chi$, for the four specific choices of $x$ and $ p $ highlighted in Fig.~\ref{fig:solnplt}. }
	\label{fig:potplot}
\end{figure}

Having obtained the operator mixing for case 1 as given in Eq.~\eqref{eq:case1-phi2}, we can substitute it back into the Lagrangian given in Eq.~\eqref{eq:lag}. The potential for the $ \chi $ field, up to $\mathcal{O}(g^2)$, assumes the following form:
{\small\begin{eqnarray}
	V(\chi) 	&=& \underbrace{\left(\frac{1}{2}M^2 +  \frac{1}{2}m^2 c^{\text{finite}}_3 \right)}_{a_1}\chi^2 + {\left(\frac{1}{4!}\lambda_\chi \right)}\chi^4  +  \mathcal{O}(g^{2n}, n>2).
\end{eqnarray}}
Note that we have ignored the term proportional to ${g^2}c^{\text{finite}}_3$ as well as the one proportional to $ g^2 K_1 $, as those are $\sim \mathcal{O}(g^4)$. Therefore, for boundedness of the potential we need $\lambda_\chi > 0$. Next if $ a_1>0$  we have the usual parabolic profile of the potential. The other interesting possibility is to have $ a_1<0 $, which implies: 

\vspace{-0.5cm}
{\small\begin{eqnarray} 
	a_1<0\; &\Rightarrow&\; ( c^{\text{finite}}_3 m^2+ M^2)  <0 \quad\Rightarrow\quad
	 c^{\text{finite}}_3 < - \frac{M^2}{m^2} \quad\Rightarrow\quad
	 c^{\text{finite}}_3 < -\frac{p}{(1-x)^2},
\end{eqnarray}}
where we have defined, $ m=m_0(1-x) $ as earlier and $ p=M^2/m_0^2 $ a dimensionless parameter. Fig.~\ref{fig:solnplt} depicts the profile of the quantity $ a_1 $ for some fixed values of $ p $ and varying $ x $. In order to plot $ a_1 $ the values of the Lagrangian parameters $g$ and $\lambda_\chi$ are kept less than unity so that perturbative expansion in terms of these parameters remains valid. The values of $x$ and $p$ are chosen in the anticipation of the change in the shape of the potential. So, for two choices of $(x,p)$, $a_1$ remains positive and the shape of the potential is parabolic and for the other two choices, it takes the shape of a wine-bottle. 
In Fig.~\ref{fig:potplot}, we have presented how the shape of $\chi$ potential changes for different benchmark points chosen from Fig.~\ref{fig:solnplt}.

\section{Conclusions and future directions}\label{sec:conclusion}

In this paper we have discussed the impact  of mass and interaction quenches, at times $t_0$ and $t_{in}$ respectively, in operator mixing and noted its consequences on the effective potential. We computed the non-equilibrium Green's functions in the closed time path formalism, having treated the mass quench exactly in the free propagator. Our free theory computation boiled down to computing correlators analogous to that of a harmonic oscillator with time-dependent frequency. The $\mathbb{Z}_2$ invariant Lagrangian which we have considered, comprises of two scalar fields $\phi$ and $\chi$. We have performed our analysis assuming two time orderings: $t_0<t_{in}\leq t<t_1<t_2$ (interaction quench occurs after mass quench) and $t_{in}< t_0< t<t_1<t_2$ (mass quench occurs after interaction quench). We have estimated the $\phi^2(t)$ operator mixing for these two cases.   The times $t_1, t_2$ are external operator times with which we find the mixings. To obtain the mixings we computed the one loop contributions to the four-point correlation function, $\langle \phi_i^2 (t)\chi_j(t_1) \chi_k(t_2)\rangle$, considering all allowed diagrams and for all $i,j,k \in \{+,-\}$.  All our answers depend explicitly on the quench parameters which include the changing mass of the $\phi$ field, the event times, $t_0, t_{in}$ and the value of field $\chi$ at $t_{in}$. In the process, we have identified the time-dependent kernels $(K_{1,2})$ and mixing coefficients $(c_3, c_3^{'})$. The kernels as well as the mixing coefficient contain similar UV divergences, which can be taken care of by employing suitable counter-terms for both cases. The mixing of the $\phi^2(t)$ operator, at leading order in the interaction, inherits the kernel's time dependence weighted by the value of the $\chi^2(t_{in})$. More interestingly, the mixing now also contains a finite part of $c_3, c_3' = c_3^{\text{finite}}$ weighted by the dynamical field $\chi^2(t)$. This finite piece as a function of $\Delta m = m - m_0$ grows negative and hence can flip the sign of the quadratic piece in the effective potential of $\chi$. Therefore, the mass quench of the $\phi$ field can trigger the possibility of late-time phase transitions in the configuration of $\chi$.  Looking ahead, there are several interesting natural directions and possible applications that we now list.

An unresolved puzzle in the context of inflation is the continuing oscillations present in the CMB data for low multipole moments \cite{Aich:2011qv}. We have seen that quenches in the action can trigger memory effects in effective potentials, leading to oscillatory behaviours in time. Quench setups are natural in our expanding universe \cite{Birrell:1982ix, Covi:2018iil}, hence it is quite envisagable that the primordial oscillations generically arise in our universe due to interacting quantum fields \textit{out of equilibrium}. 

An obvious generalization of our set-up is to consider smooth quenches characterized by an amplitude and a rate. Interesting and universal scalings are known to emerge near the breakdown of adiabaticity controlled by the quench protocol \cite{Das:2016eao}. We expect that the memory kernels will get imbued with characteristic scalings, that may indicate the presence of any phase transitions if present. Another generalization of the quench protocol is towards incorporating multiple quenches whose physics is quite different than a single quench \cite{Ghosh:2019yjh}. In particular, thermalization can occur exponentially fast and the integrable nature of a system gets washed out very quickly. 

The discussion of symmetries itself becomes complicated \textit{out of equilibrium} as the dynamics is no longer confined to the ground state\footnote{The ground state characterizes the equilibrium phase of matter.}. This is closely tied with the generation of an effective thermalization of the system, that can, in most cases, explicitly restore spontaneously broken symmetries, see however \cite{Chai:2020onq}. The connection with quenches comes via the Eigenstate Thermalization Hypothesis \cite{Srednicki:1994mfb} which posits that every finite density energy eigenstate is approximately thermal. Any \textit{out of equilibrium} scenario necessarily involves excited states, hence it is natural that an interacting system may effectively thermalize.

Finally, to draw the correct physics lessons, we need to go beyond perturbative methods. This may be possible in large $N$ theories when the system becomes exactly solvable. Recently, \cite{Chaykov:2022zro, Chaykov:2022pwd} have resummed loop contributions to the {\em in-in} correlators using the Weisskopf-Wigner method. It will be important to use this tool in our setup to find the non-perturbative results in time dependence.  

\section{Acknowledgements}
{\sf DD and BD would like to acknowledge the support provided by the Max Planck Partner Group grant MAXPLA/PHY/2018577. DD would also like to acknowledge the support provided by the MATRICS grant SERB/PHY/2020334. BD also acknowledges MHRD, Government of India for Research Fellowship. The work of JC and SUR is supported by the Science and Engineering Research Board, Government of India, under the agreement SERB/PHY/2019501
(MATRICS). SP is supported by the MHRD, Government of India, under the Prime Minister's Research Fellows (PMRF) Scheme, 2020.}


\appendix

\section{Details of loop calculations}\label{sec:loop-calc}

The contribution from the one-loop diagrams of Fig.~\ref{fig:loop-diag} ($L_1$ and $L_2$), as well as the additional diagrams ($L_1^\prime$ and $L_2^\prime$) obtained on exchanging $\vec{p}_1$ and $\vec{p}_2$, to the four-point correlation function $\langle\phi^2_+(t)\,\chi_+(t_1)\,\chi_+(t_2)\rangle_c$ can be expressed as the following integrals:

{\small\begin{eqnarray}\label{eq:L1-L1p-full}
		L_1 + L_1^\prime &=& 2\times\frac{-ig^2}{2}\int\frac{d^3\vec{k}}{(2\pi)^3}\int\limits_{t_{in}}^{\infty} d\tau\,\Big[\,G^{\phi}_{++}\left(\vec{k},t,\tau\right)\,G^{\phi}_{++}\left(\vec{P}-\vec{k},t,\tau\right)\,\times\nonumber\\
		&& \qquad\qquad \qquad\qquad G^{\chi}_{++}\left(\vec{p}_1,\tau,t_1\right)\,G^{\chi}_{++}\left(\vec{p}_2,\tau,t_2\right)\Big] \,\,+\,\, (\vec{p}_1 \leftrightarrow \vec{p}_2)\;,
\end{eqnarray}}
\vspace{-0.9cm}
{\small\begin{eqnarray}\label{eq:L2-L2p-full}
		L_2 + L_2^\prime &=& 2\times\frac{ig^2}{2}\int\frac{d^3\vec{k}}{(2\pi)^3}\int\limits_{t_{in}}^{\infty} d\tau\,\Big[\,G^{\phi}_{+-}\left(\vec{k},t,\tau\right)\,G^{\phi}_{+-}\left(\vec{P}-\vec{k},t,\tau\right)\,\times \nonumber\\
		&& \qquad\qquad \qquad\qquad G^{\chi}_{-+}\left(\vec{p}_1,\tau,t_1\right)\,G^{\chi}_{-+}\left(\vec{p}_2,\tau,t_2\right)\Big] \,\,+\,\, (\vec{p}_1 \leftrightarrow \vec{p}_2)\;.
\end{eqnarray}}
The non-trivial form of the Green's functions of the $\phi$ field, due to the sudden quench of its mass, makes it difficult to evaluate these integrals. The full expansion of the integrand, after multiplication of the terms within the Green's functions, leads to a large number of terms but the presence of $\theta$-functions, see Eqs.~\eqref{eq:gpmk} and \eqref{eq:gpp-gmm}, splits the full integral into multiple pieces with various limits of integration and one can then write,
\\
\\
\noindent \underline{\bf For case 1:}

\vspace{-0.5cm}
{\small\begin{eqnarray}\label{eq:case1-L1-L1p-split}
		L_1 + L_1^\prime &=& 2\times\frac{-ig^2}{2}\int\frac{d^3\vec{k}}{(2\pi)^3}\nonumber\\
		&&\hspace{-0.5cm}\Bigg[\int_{t_{in}}^{t}d\tau\, \frac{e^{i\Omega_{p_1}\left(\tau-t_1\right)}}{2\Omega_{p_1}} \frac{e^{i\Omega_{p_2}\left(\tau-t_2\right)}}{2\Omega_{p_2}}\, u_{in}\left(\vec{k},t\right) u_{in}\left(\vec{P}-\vec{k},t\right) u_{in}^*\left(\vec{k},\tau\right) u_{in}^*\left(\vec{P}-\vec{k},\tau\right) \nonumber\\
		&&\hspace{-0.5cm}+\int_{t}^{t_1}d\tau\, \frac{e^{i\Omega_{p_1}\left(\tau-t_1\right)}}{2\Omega_{p_1}} \frac{e^{i\Omega_{p_2}\left(\tau-t_2\right)}}{2\Omega_{p_2}}\, u_{in}\left(\vec{k},\tau\right) u_{in}\left(\vec{P}-\vec{k},\tau\right) u_{in}^*\left(\vec{k},t\right) u_{in}^*\left(\vec{P}-\vec{k},t\right) \nonumber\\ 
		&&\hspace{-0.5cm}+\int_{t_1}^{t_2}d\tau\, \frac{e^{-i\Omega_{p_1}\left(\tau-t_1\right)}}{2\Omega_{p_1}} \frac{e^{i\Omega_{p_2}\left(\tau-t_2\right)}}{2\Omega_{p_2}}\, u_{in}\left(\vec{k},\tau\right) u_{in}\left(\vec{P}-\vec{k},\tau\right) u_{in}^*\left(\vec{k},t\right) u_{in}^*\left(\vec{P}-\vec{k},t\right) \nonumber\\
		&&\hspace{-0.5cm}+\int_{t_2}^{\infty}d\tau\,\frac{e^{-i\Omega_{p_1}\left(\tau-t_1\right)}}{2\Omega_{p_1}} \frac{e^{-i\Omega_{p_2}\left(\tau-t_2\right)}}{2\Omega_{p_2}}\, u_{in}\left(\vec{k},\tau\right) u_{in}\left(\vec{P}-\vec{k},\tau\right) u_{in}^*\left(\vec{k},t\right) u_{in}^*\left(\vec{P}-\vec{k},t\right) \Bigg]\nonumber\\ 
		&& \,\,+ \,\,\left(\vec{p}_1\leftrightarrow\vec{p}_2\right) ,
\end{eqnarray}} 
\vspace{-0.8cm}
{\small\begin{eqnarray}\label{eq:case1-L2-L2p-split}
		L_2 + L_2^\prime &=& 2\times\frac{ig^2}{2}\int\frac{d^3\vec{k}}{(2\pi)^3}\nonumber\\
		&&\hspace{-0.5cm}\Bigg[\int_{t_{in}}^{\infty}d\tau\,\frac{e^{-i\Omega_{p_1}\left(\tau-t_1\right)}}{2\Omega_{p_1}} \frac{e^{-i\Omega_{p_2}\left(\tau-t_2\right)}}{2\Omega_{p_2}}\, u_{in}\left(\vec{k},\tau\right) u_{in}\left(\vec{P}-\vec{k},\tau\right) u_{in}^*\left(\vec{k},t\right) u_{in}^*\left(\vec{P}-\vec{k},t\right)  \Bigg]\nonumber\\ 
		&& \,\,+ \,\,\left(\vec{p}_1\leftrightarrow\vec{p}_2\right) .
\end{eqnarray}} 

\vspace{-0.5cm}
\noindent \underline{\bf For case 2:}
{\small\begin{eqnarray}\label{eq:case2-L1-L1p-split}
		L_1 + L_1^\prime &=& 2\times\frac{-ig^2}{2}\int\frac{d^3\vec{k}}{(2\pi)^3}\nonumber\\
		&&\hspace{-0.5cm}\Bigg[\int_{t_{in}}^{t_0}d\tau\,\frac{e^{2i\omega_{0k}\,\left(\tau-t_0\right)}}{2\,\omega_{0k}}\frac{e^{i\Omega_{p_1}\left(\tau-t_1\right)}}{2\Omega_{p_1}} \frac{e^{i\Omega_{p_2}\left(\tau-t_2\right)}}{2\Omega_{p_2}}\, u_{in}\left(\vec{k},t\right) u_{in}\left(\vec{P}-\vec{k},t\right) \nonumber\\
		&&\hspace{-0.5cm}+\int_{t_0}^{t}d\tau\,\frac{e^{i\Omega_{p_1}\left(\tau-t_1\right)}}{2\Omega_{p_1}} \frac{e^{i\Omega_{p_2}\left(\tau-t_2\right)}}{2\Omega_{p_2}}\, u_{in}\left(\vec{k},t\right) u_{in}\left(\vec{P}-\vec{k},t\right) u_{in}^*\left(\vec{k},\tau\right) u_{in}^*\left(\vec{P}-\vec{k},\tau\right) \nonumber\\ 
		&&\hspace{-0.5cm}+\int_{t}^{t_1}d\tau\,\frac{e^{i\Omega_{p_1}\left(\tau-t_1\right)}}{2\Omega_{p_1}} \frac{e^{i\Omega_{p_2}\left(\tau-t_2\right)}}{2\Omega_{p_2}}\, u_{in}\left(\vec{k},\tau\right) u_{in}\left(\vec{P}-\vec{k},\tau\right) u_{in}^*\left(\vec{k},t\right) u_{in}^*\left(\vec{P}-\vec{k},t\right) \nonumber\\
		&&\hspace{-0.5cm}+\int_{t_1}^{t_2}d\tau\,\frac{e^{-i\Omega_{p_1}\left(\tau-t_1\right)}}{2\Omega_{p_1}} \frac{e^{i\Omega_{p_2}\left(\tau-t_2\right)}}{2\Omega_{p_2}}\, u_{in}\left(\vec{k},\tau\right) u_{in}\left(\vec{P}-\vec{k},\tau\right) u_{in}^*\left(\vec{k},t\right) u_{in}^*\left(\vec{P}-\vec{k},t\right) \nonumber\\
		&&\hspace{-0.5cm}+\int_{t_2}^{\infty}d\tau\,\frac{e^{-i\Omega_{p_1}\left(\tau-t_1\right)}}{2\Omega_{p_1}} \frac{e^{-i\Omega_{p_2}\left(\tau-t_2\right)}}{2\Omega_{p_2}}\, u_{in}\left(\vec{k},\tau\right) u_{in}\left(\vec{P}-\vec{k},\tau\right) u_{in}^*\left(\vec{k},t\right) u_{in}^*\left(\vec{P}-\vec{k},t\right) \Bigg]\nonumber\\ 
		&& \,\,+ \,\,\left(\vec{p}_1\leftrightarrow\vec{p}_2\right) ,
\end{eqnarray}} 
\vspace{-0.8cm}
{\small\begin{eqnarray}\label{eq:case2-L2-L2p-split}
		L_2 + L_2^\prime &=& 2\times\frac{ig^2}{2}\int\frac{d^3\vec{k}}{(2\pi)^3}\nonumber\\
		&&\hspace{-0.5cm}\Bigg[\int_{t_{in}}^{t_0}d\tau\,\frac{e^{-2i\omega_{0k} \left(\tau-t_0\right)}}{2\,\omega_{0k}} \frac{e^{-i\Omega_{p_1}\left(\tau-t_1\right)}}{2\Omega_{p_1}} \frac{e^{-i\Omega_{p_2}\left(\tau-t_2\right)}}{2\Omega_{p_2}}\, u_{in}^*\left(\vec{k},t\right) u_{in}^*\left(\vec{P}-\vec{k},t\right) \nonumber\\ 
		&&\hspace{-0.5cm}+\int_{t_0}^{t_\infty}d\tau\,\frac{e^{-i\Omega_{p_1}\left(\tau-t_1\right)}}{2\Omega_{p_1}}\frac{e^{-i\Omega_{p_2}\left(\tau-t_2\right)}}{2\Omega_{p_2}}\, u_{in}\left(\vec{k},\tau\right) u_{in}\left(\vec{P}-\vec{k},\tau\right) u_{in}^*\left(\vec{k},t\right) u_{in}^*\left(\vec{P}-\vec{k},t\right) \Bigg] \nonumber\\
		&& \,\,+ \,\,\left(\vec{p}_1\leftrightarrow\vec{p}_2\right) .
\end{eqnarray}} 
For both case 1 and case 2, the total contribution to the correlation function is obtained after summing up $L_1 + L_1^\prime + L_2 + L_2^\prime$. Eqs.~\eqref{eq:case1-L1-L1p-split} - \eqref{eq:case2-L2-L2p-split} are further simplified by appropriately substituting for $u_{in}$, $u_{in}^*$ using Eq.~\eqref{eq:u-in-kt} and by working in a large-$\vec{k}$ limit which implies: $(\vec{P} -\vec{k})^2 + m^2 \approx \vec{k}^2 + m^2$. Subsequently, the integration over $\tau$ can be done using computational tools such as \texttt{Mathematica} \cite{Mathematica}. After the integration, the result can be further filtered by keeping only the leading and sub-leading powers of $\vec{k}$ in the numerator as well as the denominator. Finally, ignoring all terms except the ones of the order $1/|\vec{k}|^3$, we can identify four-point correlation functions of $\chi$ along with their time-dependent coefficients. This procedure ultimately leads to the contents of Eqs.~\eqref{eq:case1-4pt-chi-corr} and \eqref{eq:case2-4pt-chi-corr} for case 1 and case 2 respectively.

\section{Series expansion of the kernels for small mass quench}\label{sec:kernel-series-expansion}

Expansion of the post-quench mass of the $\phi$ field ($m$) around the pre-quench mass ($m_0$) as $m = m_0 + \delta m$, assuming a small $\delta m$, leads to the following relation between $\omega_k$ and $\omega_{0k}$:
$$\omega_k = \omega_{0 k} + \delta \omega_k \hspace{0.5cm} \text{with} \hspace{0.5cm}  \delta \omega_k =\frac{m_0}{\omega_{0 k}} \delta m.$$

Based on this, the kernels $K_1$ and $c_3$, defined in Eqs.~\eqref{K1} and \eqref{c3} respectively, can also be expanded, up to linear order in $\delta m$ as:

{\small\begin{eqnarray}\label{eq:K1-expanion}
		&&\hspace{-1.5cm}K_1(t-t_{in} ;\,  t_{in} - t_0) = \frac{g^2}{4}\int \frac{d^3\vec{k}}{(2 \pi)^3} \frac{\cos 2\omega_{0 k}(t-t_{in})}{\omega^3_{0 k}}\nonumber\\ 
		&-& \frac{g^2 m_0}{4}\, \delta m \Bigg[3 \int \frac{d^3\vec{k}}{(2 \pi)^3} \frac{\cos 2\omega_{0 k} (t-t_{in})}{\omega^5_{0 k}} + 2 \int \frac{d^3\vec{k}}{(2 \pi)^3} (t-t_{in}) \frac{ \sin 2\omega_{0 k} (t - t_{in})}{\omega^4_{0 k}}   \Bigg] \nonumber\\
		& = & \cfrac{g^2}{32\pi}\, G^{2,0}_{1,3}\,\Big(z^2\,\Big|^{\;\;,3/2}_{0,0,1/2}\Big) -\frac{g^2 \delta m}{32 \pi m_0}  \Bigg[3 \, G^{2,0}_{1,3}\,\Big(z^2\,\Big|^{\;\;,5/2}_{1,0,1/2}\Big) -z \cfrac{d}{dz}\Big[G^{2,0}_{1,3}\,\Big(z^2\,\Big|^{\;\;,5/2}_{1,0,1/2}\Big)\Big] \Bigg], 
\end{eqnarray}}
\vspace{-0.6cm}
{\small\begin{eqnarray}\label{eq:c3-expanion}
		c_3(t, t_0)  &=&  -\frac{g^2}{4}  \int \frac{d^3\vec{k}}{(2 \pi)^3} \frac{1}{\omega^3_{0 k}} + \frac{3m_0\, g^2}{4} \int \frac{d^3\vec{k}}{(2 \pi)^3} \, \frac{1}{\omega^5_{0 k}}  \delta m \nonumber\\
		&=& -\frac{g^2}{8\pi^2}\left[-\frac{\Lambda}{\sqrt{\Lambda^2 + m_0^2}} + \log \left(\frac{\Lambda}{m_0}+\sqrt{1+\frac{\Lambda^2}{m_0^2}}\right) \right] + \frac{ g^2}{8\pi^2 m_0}  \delta m,\hspace{.5cm}
\end{eqnarray}}
where, $ z=m_0(t-t_{in}) $, with $m_0$ assumed to be a constant and $\Lambda$ is the UV cut-off. 
It must be noted that the zeroth order term is the same as what was reported in \cite{Dresti:2013kya}. 

For case 2, the kernels $K_2$ and $c_3^\prime$, see Eqs.~\eqref{K2} and \eqref{c3p}, can similarly be expanded:

\vspace{-0.4cm}
{\small\begin{eqnarray}\label{eq:K2-expanion}
		&&\hspace{-1.5cm}K_2(t-t_{in} ;\,  t_{in} - t_0) =\frac{g^2}{4}  \int \frac{d^3\vec{k}}{(2 \pi)^3} \frac{\cos 2\omega_{0 k}(t-t_{in})}{\omega^3_{0 k}} \nonumber\\ 
		&-&\frac{ m_0\, \delta m \, g^2}{4} \Bigg[ \int \frac{d^3\vec{k}}{(2 \pi)^3} \frac{\cos 2\omega_{0 k} (t-t_{in})}{\omega^5_{0 k}} + \int \frac{d^3\vec{k}}{(2 \pi)^3} 2(t - t_0) \frac{ \sin 2\omega_{0 k} (t - t_{in})}{\omega^4_{0 k}}   \Bigg] \nonumber\\
		&=& \cfrac{g^2}{32\pi}\, G^{2,0}_{1,3}\,\Big(z^2\,\Big|^{\;\;,3/2}_{0,0,1/2}\Big) -\frac{\delta m \, g^2}{32 \pi m_0} \Bigg[G^{2,0}_{1,3}\,\Big(z^2\,\Big|^{\;\;,5/2}_{1,0,1/2}\Big) -z\cfrac{d}{dz}\Big[G^{2,0}_{1,3}\,\Big(z^2\,\Big|^{\;\;,5/2}_{1,0,1/2}\Big)\Big]  \nonumber\\
		& & \hspace{5cm}+(t_{in}-t_0)\cfrac{d}{dz}\Big[G^{2,0}_{1,3}\,\Big(z^2\,\Big|^{\;\;,5/2}_{1,0,1/2}\Big)\Big] \Bigg] ,
\end{eqnarray}}
{\small\begin{eqnarray}\label{eq:c3p-expanion}
		c_3^\prime(t, t_0)  &=&  - \frac{g^2}{4} \int \frac{d^3\vec{k}}{(2 \pi)^3} \frac{1}{\omega^3_{0 k}} + \frac{3m_0\, g^2}{4}  \int \frac{d^3\vec{k}}{(2 \pi)^3} \, \frac{1}{\omega^5_{0 k}}  \delta m \nonumber\\
		&=& -\frac{g^2}{8\pi^2}\left[-\frac{\Lambda}{\sqrt{\Lambda^2 + m_0^2}} + \log \left(\frac{\Lambda}{m_0}+\sqrt{1+\frac{\Lambda^2}{m_0^2}}\right) \right] + \frac{ g^2}{8\pi^2 m_0} \delta m\;. \hspace{.5cm}
\end{eqnarray}}
\noindent
The $\vec{k}$-integrals within each of these expressions can be rewritten in terms of Meijer-G functions. In Table~\ref{tab:meijer-g}, we have listed the relevant Meijer-G functions and their asymptotic functional forms.

\begin{table}[h]
	\centering
	\renewcommand{\arraystretch}{3}
	{\small\begin{tabular}{|c|c|c|}
			\hline
			\textsf{Meijer-G function}&
			\textsf{Large $z$ limit}&
			\textsf{Small $z$-limit}\\
			\hline

			$G^{2,0}_{1,3}\,\Big(z^2\,\Big|^{\;\;,3/2}_{0,0,1/2}\Big)$&
			$-\cfrac{z^{-3/2}}{\sqrt{2\pi}}\,\left(\cos\left(2z\right)+\sin\left(2z\right)\right)$&
			$-\cfrac{4}{\pi}\left(1+\gamma_{_E}+\log(z)\right)$\\

			$G^{2,0}_{1,3}\,\Big(z^2\,\Big|^{\;\;,5/2}_{1,0,1/2}\Big)$&
			$-\cfrac{z^{-3/2}}{\sqrt{2\pi}}\,\left(\cos\left(2z\right)+\sin\left(2z\right)\right)$&
			$\cfrac{4}{3\pi }$\\

			$\cfrac{d}{dz}\Big[G^{2,0}_{1,3}\,\Big(z^2\,\Big|^{\;\;,5/2}_{1,0,1/2}\Big)\Big]$&
			$-2\cfrac{z^{-3/2}}{\sqrt{2\pi}}\,\left(\cos\left(2z\right)-\sin\left(2z\right)\right)$&
			$\cfrac{16 z}{\pi }\,(\gamma_{_E} + \log(z))$\\
			
			\hline
	\end{tabular}}
	\caption{\sf Meijer-G functions corresponding to the integrals present in the Kernel definitions, along with their asymptotic limits. For the limiting cases, we have only reported the first non-zero term of the series expansion. Here, $ z=m_0(t-t_{in})$. }
	\label{tab:meijer-g}
\end{table}

\providecommand{\href}[2]{#2}
\addcontentsline*{toc}{section}{}
\bibliographystyle{jhep}
\bibliography{refs}

\end{document}